%% file: paper.tex
\documentclass [a4paper,fleqn] {article}
\usepackage{graphics,epic,eepic,a4wide,epsfig}
\usepackage{latexsym}

\newtheorem{theorem}{Theorem}[section]
\newtheorem{lemma}[theorem]{Lemma}
\newtheorem{corollary}[theorem]{Corollary}
\newcommand{\pf}{\par\noindent{\bf Proof:}~}
\newcommand{\qed}{\hfill{\rule{3mm}{3mm}}\medskip}
\newenvironment{proof}{\pf}{\qed}
\newtheorem{proposition}[theorem]{Proposition}
\newenvironment{example}
{\refstepcounter{theorem}{\vspace{2ex}\par\noindent \bf
Example~\thetheorem~}}{\vspace{2ex}\par}

\input{preamble}

\begin{document}

\title{A Theory of Normed Simulations\thanks{A preliminary version of this
paper appeared as Sections 1 and 2 in \cite{GV98a}.}}

\author{David Griffioen\thanks{Supported by the Netherlands Organization for
Scientific Research (NWO) under contract SION 612-316-125.
Current e-mail address: {\tt griffioen42@zonnet.nl}.}\\
%\and
Frits Vaandrager\thanks{E-mail: {\tt fvaan@cs.kun.nl}.}\\
Nijmeegs Instituut voor Informatica en Informatiekunde, University of Nijmegen\\
P.O.\ Box 9010, 6500 GL Nijmegen, The Netherlands
}

\date{}

\maketitle

\begin{abstract}
In existing simulation proof techniques, a single step in a lower-level
specification may be simulated by an extended execution fragment in a
higher-level one.
As a result, it is cumbersome to mechanize these techniques using
general purpose theorem provers.
Moreover, it is undecidable whether a given relation is a simulation,
even if tautology checking is decidable for the underlying specification logic.
This paper studies various types of {\em normed simulations}.
In a normed simulation, each step in a lower-level specification can be
simulated by at most one step in the higher-level one, for any related
pair of states.
In earlier work we demonstrated that normed simulations are quite useful
as a vehicle for the formalization of refinement proofs via theorem provers.
Here we show that normed simulations also have
pleasant theoretical properties:
(1) under some reasonable assumptions, it is decidable whether a
given relation is a normed forward simulation,
provided tautology checking is decidable for the underlying logic;
(2) at the semantic level, normed forward and backward
simulations together form a complete proof method for establishing behavior
inclusion, provided that the higher-level specification has finite
invisible nondeterminism.

\noindent
{\bf AMS Subject Classification (1991):}
%Specification and verification of programs:
68Q60,
%Automata theory:
68Q68.

\noindent
{\bf CR Subject Classification (1991):}
%Models of Computation, automata:
F.1.1,
%Specifying, Verifying and Reasoning about programs:
F.3.1.

\noindent
{\bf Keywords \& Phrases:}
Computer aided verification,
normed simulations,
automata,
refinement mappings,
forward simulations,
backward simulations,
history variables,
prophecy variables.
\end{abstract}

\pagebreak

\tableofcontents

\pagebreak

\input{intro}
\input{normed}

\subsection*{Acknowledgement}
We thank Mari\"elle Stoelinga for spotting a mistake in an earlier version
of this paper, and Jan Willem Klop for discussions that led us to
Theorem~\ref{Tm:history po}.

\bibliographystyle{alpha}
\bibliography{abbreviations,dbase}

\end{document}

%% file: preamble.tex
\newcommand{\ms}[1]{\ifmmode%
\mathord{\mathcode`-="702D\it #1\mathcode`\-="2200}\else%
$\mathord{\mathcode`-="702D\it #1\mathcode`\-="2200}$\fi}

\newcommand{\deq}{\mathrel{\stackrel{\scriptscriptstyle\Delta}{=}}}
\newcommand{\implies}{\mathrel{\Rightarrow}}
\newcommand{\Iff}{\mathrel{\Leftrightarrow}}

\newcommand{\nat}{{\sf N}}

\newcommand{\first}[1]{\ms{first{(#1)}}}
\newcommand{\last}[1]{\ms{last{(#1)}}}
\newcommand{\Index}[1]{\ms{index{(#1)}}}

\newcommand{\inverse}[1]{#1^{-1}}
\newcommand{\domain}[1]{\ms{domain{(#1)}}}

\newcommand{\states}[1]{\ms{states{(#1)}}}
\newcommand{\start}[1]{\ms{start{(#1)}}}
\newcommand{\acts}[1]{\ms{acts{(#1)}}}
\newcommand{\steps}[1]{\ms{steps{(#1)}}}
\newcommand{\ext}[1]{\ms{ext{(#1)}}}
\newcommand{\arrow}[1]{
                  \raisebox{0ex}[2.1ex][0ex]{
                  $\stackrel{\!\!\!#1}{\raisebox{0ex}[0.6ex][0ex]{$
                  \negthinspace-\!\!\!\rightarrow\:$}}$}}
\newcommand{\sarrow}[2]{
                  \raisebox{0ex}[2.1ex][0ex]{
                  $\stackrel{\!\!\!#1\:}{\raisebox{0ex}[0.6ex][0ex]{$
                  \negthinspace-\!\!\!\rightarrow_{#2}\;$}}$}}
\newcommand{\darrow}[1]{
                  \raisebox{0ex}[2.1ex][0ex]{
                  $\stackrel{#1 ~~}{\raisebox{0ex}[0.6ex][0ex]{$
                  \negthinspace=\!\!\!\Rightarrow\:$}}$}}
\newcommand{\sdarrow}[2]{
                  \raisebox{0ex}[2.1ex][0ex]{
                  $\stackrel{#1}{\raisebox{0ex}[0.6ex][0ex]
		  {$\negthinspace=\!\!\!\Rightarrow$}   }_{#2}$}}

\newcommand{\fexecs}[1]{\ms{execs}^{\ast}(#1)}
\newcommand{\execs}[1]{\ms{execs}(#1)}

\newcommand{\trace}[1]{{\it trace}(#1)}

\newcommand{\ftraces}[1]{{\it traces}^{\ast}(#1)}
\newcommand{\traces}[1]{{\it traces}(#1)}

\newcommand{\after}[1]{{\it after}(#1)}
\newcommand{\past}[1]{{\it past}(#1)}

\newcommand{\pre}[1]{\leq_{{\rm #1}}}
\newcommand{\preft}{\pre{\ast T}}
\newcommand{\pret}{\pre{T}}
\newcommand{\prer}{\pre{R}}
\newcommand{\pref}{\pre{F}}
\newcommand{\preb}{\pre{B}}
\newcommand{\preifb}{\pre{iB}}
\newcommand{\preh}{\pre{H}}
\newcommand{\prep}{\pre{P}}
\newcommand{\preifp}{\pre{iP}}

\newcommand{\unfold}[1]{\ms{unfold{(#1)}}}
\newcommand{\superp}[3]{\ms{sup{(#1,#3,#2)}}}

%% file: intro.tex
\section{Introduction}

Simulation relations and refinement functions are widely used to prove that
a lower-level specification of a reactive system correctly
implements a higher-level one \cite{Jo94,Ly96,dRE98}.
Proving soundness and completeness of proof rules for simulation and
refinement has attracted the attention of many researchers in the past two or
three  decades \cite{Mi71,Lam83,Jo85,LT87,Sta88,KS89,KS93,Jo90,Jo91,AL91,LV95}.
The usefulness of all these proof methods was demonstrated by their proposers,
who applied them to often highly nontrivial case studies.  However, all these
refinement/simulation proofs were done manually, and they were typically quite
long and tedious.
The field has come to realize that if we want to scale up these methods to
larger examples, it really matters that the semantical analysis can be carried
out with the help of a software tool that requires little or no human
intervention.  This led Wolper \cite{Wol97}
to propose the following criterion for ``formal'' methods
\begin{quote}
{\bf Criterion of Semantical Computational Support:}
{\it A formal method provides semantical computational support
of it allows software tools for checking semantical properties
of specifications.}
\end{quote}
Several incomplete refinement/simulation proof rules have been mechanized
successfully \cite{HSV94,NipS94,DGRV00}.
A mechanization of a complete set of simulation rules is reported by
Sogaard-Andersen et al.\ \cite{SGGLP93}, but in this approach the verification
process is highly interactive and it does not satisfy Wolper's criterion of
semantical computational support.
In fact, we believe it will be difficult to efficiently mechanize any of the
above mentioned complete proof methods using a general purpose theorem prover:
too much user interaction will be required.
Earlier \cite{GV98a,Gri00}[Chapter 6], we proposed a proof method based on
{\em normed simulations} and showed that it can be mechanized efficiently
using PVS.
In the present paper we study the theoretical properties of normed simulations.
In particular, we establish that normed forward and backward simulations
together form a complete proof method for establishing behavior inclusion.
Before we discuss the technical contributions of this paper in more detail,
we first describe the problem that arises in the mechanization of
existing complete proof methods, and how this can be solved using
normed simulations.

\begin{figure}[ht]
\begin{center}
\epsfig{file=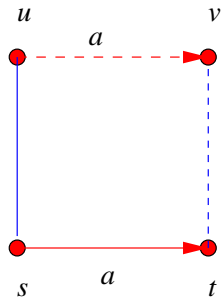}
\caption{\label{Fe:transfer1}Transfer condition (\ref{transfer1}).}
\end{center}
\end{figure}

%Often the lower-level specification is referred to as the {\em implementation}
%and the higher-level specification simply as the {\em specification},
%but this terminology is somewhat confusing since an implementation may
%in turn serve as specification later on in the design process.
Technically, a {\em simulation} (or {\em refinement}) is a relation
(or function) $R$ between the states of a lower-level specification $A$
and a higher-level specification $B$, that satisfies a condition like
\begin{eqnarray}
\label{transfer1}
(s,u) \in R \:\wedge\: s \sarrow{a}{A} t & \implies &
\exists v : u \sarrow{a}{B} v \: \wedge \: (t,v) \in R
\end{eqnarray}
(If lower-level state $s$ and higher-level state $u$ are related,
and in $A$ there is a transition from $s$ to $t$,
then there is a matching transition in $B$ from $u$ to a state $v$
that relates to $t$; see also Figure~\ref{Fe:transfer1}.)
The existence of a simulation implies that any behavior of $A$ can also be
exhibited by $B$.

\begin{figure}[ht]
\begin{center}
\epsfig{file=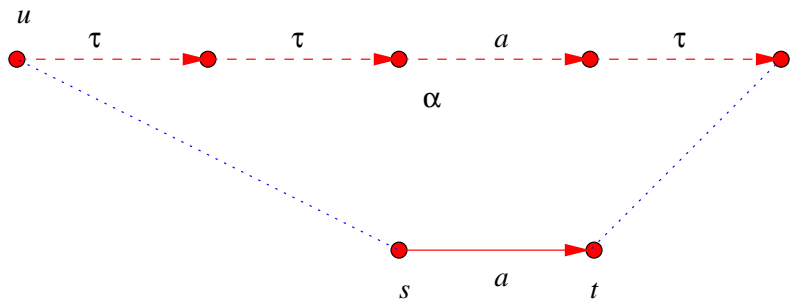}
\caption{\label{Fe:transfer2}Transfer condition (\ref{transfer2}).}
\end{center}
\end{figure}

The main reason why simulations are useful is that they reduce {\em global}
reasoning about behaviors and executions to {\em local} reasoning about states
and transitions.
However, to the best of our knowledge, all complete simulation proof methods
that appear in the literature fall back on some form of global reasoning
in the case of specifications containing internal (or stuttering) transitions.
The usual transfer condition for {\em forward simulations} \cite{LV95},
for instance, says
\begin{eqnarray}
\label{transfer2}
(s,u) \in R \:\wedge\: s \sarrow{a}{A} t & \implies &
\exists \mbox{ execution fragment } \alpha : \first{\alpha} = u \\
& & ~ \wedge\: \trace{\alpha} = \trace{a} \:\wedge\:
(t, \last{\alpha}) \in R  \nonumber
\end{eqnarray}
(Each lower-level transition can be simulated by a sequence of higher-level
transitions which, apart from the action that has to be matched, may also
contain an arbitrary number of internal ``$\tau$'' transitions;
see also Figure~\ref{Fe:transfer2}.)
Thus the research program to reduce global reasoning to local reasoning
has not been carried out to its completion.
In manual proofs of simulation relations, this is usually not a problem:
in practice lower-level transitions
are typically matched by at most one higher-level transition;
moreover humans tend to be quite good in reasoning about sequences,
and move effortlessly from transitions to executions and back.
In contrast, it turns out to be rather cumbersome to formalize arguments
involving sequences using existing theorem provers \cite{DGM97}.
In fact, in several papers in which formalizations of simulation proofs
are described, the authors only consider a restricted type of simulation
in which each lower-level transition is matched by at most one higher-level
transition \cite{HSV94,NipS94,DGRV00}.
However, there are many examples of situations where these restricted
types of simulations cannot be applied.
In approaches where the full transfer condition
(\ref{transfer2}) is formalized \cite{SGGLP93},
the user has to supply the simulating
execution fragments $\alpha$ to the prover explicitly,
which makes the verification process highly interactive.
Jonsson \cite{Jo90} presents a variant of the completeness theorem
of Abadi and Lamport \cite{AL91} in terms of certain forward and backward
simulations in which lower-level transitions are matched by at most
one higher-level transition.  However, his completeness result is only
partial in the sense that he requires that the higher-level automaton
contains no non-trivial $\tau$-steps.  In our view this restriction is
problematic, especially in a stepwise refinement approach where the higher-level
specification in one design step may be the lower-level specification from a
previous design step.
All the complications that we address in our paper are due to the possible
presence of internal actions in the higher-level automaton.

In this paper, we study a simulation proof method which remedies the
above problems.  The idea is to define a function $n$
that assigns a norm $n(s \arrow{a} t,u)$, in some well-founded domain,
to each pair of a transition in $A$ and a state of $B$.
If $u$ has to simulate transition $s \arrow{a} t$ then it may either do nothing
(if $a$ is internal and $t$ is related to $u$),
or it may do a matching $a$-transition, or it may perform an internal
transition $u \arrow{b} v$ such that the norm decreases, i.e.,
\begin{eqnarray*}
n(s \arrow{a} t, v) & < & n(s \arrow{a} t,u).
\end{eqnarray*}
We establish that {\em normed forward simulations} and {\em normed backward
simulations} together constitute a complete proof method for establishing trace
inclusion.  In addition we show how {\em history} and {\em prophecy relations}
(which are closely related to history and prophecy variables \cite{AL91})
can be enriched with a norm function, to obtain another complete proof method
in combination with a simple notion of refinement mapping.

The preorders generated by normed forward simulations are strictly finer than
the preorders induced by Lynch and Vaandrager's forward simulations \cite{LV95}.
In fact, we will characterize normed forward simulations in terms of
{\em branching forward simulations} \cite{GW96}, and present a
similar characterization for the backward case.
It is possible to come up with a variant of normed forward simulation that
induces the same preorder as forward simulations, but technically this is
somewhat more involved \cite{Gri00}[Section 6.5.10].

When proving invariance properties of programs, one is faced with two problems.
The first problem is related to the necessity of proving tautologies of the
assertion logic, whereas the second manifests in the need of finding
sufficiently strong invariants.
In order to address the first problem, powerful decision procedures have
been incorporated in theorem provers such as PVS \cite{ORSH95}.
If tautology checking is decidable then it is decidable whether a given
state predicate is valid for the initial states and preserved by all
transitions.  The task of finding such a predicate, i.e.\ solving the second
problem, is in most cases still the responsibility of the user, even though
some very powerful heuristics have been devised to support and automate
the search \cite{BLS96,MBSU98,LBBO01,SALoverview00:LFM}.
Analogously, if specifications $A$ and $B$,
a conjectured forward simulation relation $R$
and norm function $n$ can all be expressed within a decidable assertion logic,
and if the specification of $B$ only contains a
finite number of deterministic transition predicates,
then it is decidable whether the pair $(R,n)$ is a normed forward simulation.
This result, which does not hold for earlier approaches such as \cite{LV95},
is a distinct advantage of normed forward simulations.

The idea of using norm functions to prove simulation relations was also
developed by Groote and Springintveld \cite{GS95}, who used it to prove
branching bisimilarity in the context of the process algebra $\mu$CRL.
However, their norm function is defined on the states of $B$ only
and does not involve the transitions of $A$.
As a consequence, their method does not always apply to diverging processes.
Norm functions very similar to ours were also studied by Namjoshi \cite{Nam97}.
He uses them to obtain a characterization of the stuttering bisimulation
of Browne et al.\  \cite{BCG88}, which is the equivalent of branching bisimulation in a
setting where states rather than actions are labeled \cite{DV95}.
Neither Groote and Springintveld \cite{GS95}, nor Namjoshi \cite{Nam97} address
effectiveness issues.
Although we present normed simulations in a setting of labeled transition
systems, it should not be difficult to transfer our results to
a process algebraic setting such as that of Groote and Springintveld
\cite{GS95} or a state based setting such as Namjoshi's \cite{Nam97}.
Inspired by our approach, norm functions have been used by Baier
and Stoelinga \cite{BS00} to
define a new bisimulation equivalence for probabilistic systems.

In this paper, we only present maximally simple examples to illustrate
the various definitions and results.
Earlier \cite{GV98a,Gri00}[Chapter 6], we used normed simulations in a
substantial case study, namely the verification of the leader
election protocol that is part of the IEEE 1394 ``Firewire'' standard.
This verification has been mechanically checked using PVS.\footnote{Actually,
we discovered the notion of a normed simulation while formalizing the
correctness proof of this leader election protocol.}

In the presentation of our results, we will closely follow
Lynch and Vaandrager \cite{LV95} and stick to their notations.
In fact, our aim will be (amongst others) to derive analogous results to
theirs, only for different types of simulations.
However, we decided not to present normed versions of their forward-backward
and backward-forward simulations of, since these simulations
have thus far not been used in practice and technically this would bring
nothing new.
Apart from the notion of a norm function, a major technical innovation in
the present paper is a new, simple definition of execution correspondence
\cite{GSSL93,SLL93},
and the systematic use of this concept in the technical development.
Although here we only address simulation proof techniques for
establishing safety, we expect that based on the execution
correpondence lemma's that we prove it will be easy to generalize
our results to a setting with liveness properties.
We leave it as a topic for future research to substantiate this claim.

%% file: normed.tex
\section{Preliminaries}

In this section, we briefly recall some basic concurrency theory definitions
\cite{LV95}.
An {\em automaton} (or {\em labeled transition system}) $A$ consists of
\begin{itemize}
\item
a (possibly infinite) set $\states{A}$ of states,
\item
a nonempty set $\start{A}\subseteq\states{A}$ of start states,
\item
a set $\acts{A}$ of actions that includes the internal (or stuttering) action
$\tau$, and
\item
a set $\steps{A} \subseteq\states{A}\times\acts{A}\times\states{A}$ of steps.
\end{itemize}
Write $s \sarrow{a}{A} t$ as a shorthand for $(s,a,t) \in\steps{A}$.
We let $\ext{A}$, the {\em external actions}, denote $\acts{A} - \{ \tau \}$.
An {\em execution fragment} of $A$ is a finite or infinite
alternating sequence, $s_0 a_1 s_1 a_2 s_2 \cdots$, of states
and actions of $A$, beginning with a state, and if it is finite
also ending with a state, such that for all $i>0$, $s_{i-1} \arrow{a_i} s_i$.
An {\em execution} of $A$ is an execution fragment that begins with a
start state.
We denote by $\fexecs{A}$ and $\execs{A}$ the sets of finite
and  all executions of $A$, respectively.
A state $s$ of $A$ is {\em reachable} if $s$ occurs as the last state
in some finite execution $\alpha$ of $A$.
In this case we write ${\it reachable}(A,s)$.
Also, we write ${\it reachable}(A)$ for the set of reachable states of $A$.

The {\em trace} of an execution fragment $\alpha$, notation $\trace{\alpha}$,
is the subsequence of non-$\tau$ actions occurring in $\alpha$.
A finite or infinite sequence $\beta$ of external
actions is a {\em trace} of
$A$ if $A$ has an execution $\alpha$ with $\beta = \trace{\alpha}$.
Write $\ftraces{A}$ and $\traces{A}$ for the sets of
finite and all traces of $A$, respectively.
Write $A \preft B$ if $\ftraces{A} \subseteq \ftraces{B}$,
and $A \pret B$ if $\traces{A} \subseteq \traces{B}$.

Suppose $A$ is an automaton, $s$ and $t$ are states of $A$,
and $\beta$ is a finite sequence over $\ext{A}$.
We say that $(s,\beta,t)$ is a {\em move} of $A$,
and write $s \sdarrow{\beta}{A} t$, or just
$s \darrow{\beta} t$ when $A$ is clear,
if $A$ has a finite execution fragment $\alpha$ that starts in
$s$, has trace $\beta$ and ends in $t$.

Three restricted kinds of automata play an important role in this paper:
\begin{enumerate}
\item
$A$ is {\em deterministic} if $|\start{A}| = 1$,
and for any state $s$ and any finite sequence $\beta$ over $\ext{A}$,
there is at most one state $t$ such that $s \darrow{\beta} t$.
A deterministic automaton is characterized uniquely by the properties
that $|\start{A}| = 1$,
every $\tau$-step is of the form $(s, \tau , s)$ for some $s$, and
for each state $s$ and each action $a$ there is at most one
state $t$ such that $s \sarrow{a}{A} t$.
\item
$A$ has {\em finite invisible nondeterminism (fin)} if
$\start{A}$ is finite, and
for any state $s$ and any finite sequence $\beta$ over $\ext{A}$,
there are only finitely many states $t$ such that $s \sdarrow{\beta}{A} t$.
\item
$A$ is a {\em forest} if, for each state $s$ of $A$, there is
exactly one execution that leads to $s$.
A forest is characterized uniquely by the property that
all states of $A$ are reachable,
start states have no incoming steps, and
each of the other states has exactly one incoming step.
\end{enumerate}
The relation $\after{A}$ consists of the pairs $( \beta ,s)$ for which there is
a finite execution of $A$ with trace $\beta$ and last state $s$:
\[
\after{A}\deq
\{ ( \beta , s) \mid \exists\alpha\in\fexecs{A} :
\trace{\alpha} = \beta \mbox{ and } \last{\alpha} = s \}.
\]
(Here ${\it last}$ denotes the function that returns the last element
of a finite, nonempty sequence.)
We also define $\past{A}$ to be the inverse of $\after{A}$,
$\past{A}\deq\inverse{\after{A}}$;
this relates a state $s$ of $A$ to
the traces of finite executions of $A$ that lead to $s$.

The following elementary lemma by Lynch and Vaandrager \cite{LV95} states
that for the restricted kinds of automata defined above, the relations
${\it after}$ and ${\it past}$ satisfy certain nice properties.

\begin{lemma}
\label{La:beforeafter}
\mbox{ }
\begin{enumerate}
\item
If $A$ is deterministic then
$\after{A}$ is a function from $\ftraces{A}$ to $\states{A}$.
\item
If $A$ has fin then $\after{A}$ is {\em image-finite}, i.e., each trace in
the domain of $\after{A}$ is only related to finitely many states.
\item
If $A$ is a forest then $\past{A}$ is a function from $\states{A}$ to
$\ftraces{A}$.
\end{enumerate}
\end{lemma}

\section{Step Refinements and Execution Correspondence}
\label{StepRefinements}

In this section, we present {\em step refinements},
the simplest notion of simulation that we consider in this paper.
In order to prove soundness of step refinements, we also introduce the
auxiliary notion of {\em execution correspondence}.
This notion plays a key role in this paper; the technical lemmas that we
prove in this section will also be used repeatedly in subsequent sections.

\subsection{Step Refinements}
\label{Dn:StepRefinements}
Let $A$ and $B$ be automata.
A {\em step refinement} from $A$ to $B$ is a partial function $r$ from
$\states{A}$ to $\states{B}$ that satisfies the following two conditions:
\begin{enumerate}
\item
If $s \in \start{A}$ then $s \in\domain{r}$ and $r(s) \in \start{B}$.
\item
If $s \sarrow{a}{A} t \:\wedge\: s \in\domain{r}$ then $t \in\domain{r}$ and
\begin{enumerate}
\item
$r(s)=r(t) \: \wedge \: a = \tau$, or
\item
$r(s) \sarrow{a}{B} r(t)$.
\end{enumerate}
\end{enumerate}
Note that, by a trivial inductive argument, the set of states for
which $r$ is defined contains all the reachable states of $A$ (and is
thus an {\em invariant} of this automaton).
We write $A \prer B$ if there exists a step refinement from $A$ to $B$.

As far as we know, the notion of step refinements was first proposed
by Nipkow and Slind \cite{NipS94}.
However, if we insist on the presence of stuttering steps
$s \arrow{\tau} s$ for each state $s$ (a common assumption in
models of reactive systems) then clause (2a) in the above definition
becomes superfluous and the notion of a step refinement reduces to
that of a homomorphism between reachable subautomata \cite{Gin68}.
Step refinements are slightly more restrictive than the
{\em possibility mappings} of Lynch and Tuttle \cite{LT87}
(called {\em weak refinements} by Lynch and Vaandrager \cite{LV95}).
In the case of a possibility mapping each (reachable)
step of $A$ may be matched by a sequence of steps in $B$ with the same trace.
This means that in the above definition condition (2) is replaced by:
\begin{enumerate}
\item[2.]
If $s \sarrow{a}{A} t \:\wedge\: s \in\domain{r}$ then $t \in\domain{r}$ and
$B$ has an execution fragment $\alpha$ with
$\first{\alpha} = r(s)$, $\trace{\alpha} = \trace{a}$ and
$\last{\alpha} = r(t)$.
\end{enumerate}
Observe that, unlike step refinements, possibility mappings do not reduce
global reasoning to local reasoning.

\begin{example}
Figure~\ref{Fe:refinement} illustrates the notion of a
step refinement.
\begin{figure}[ht]
\begin{center}
\epsfig{file=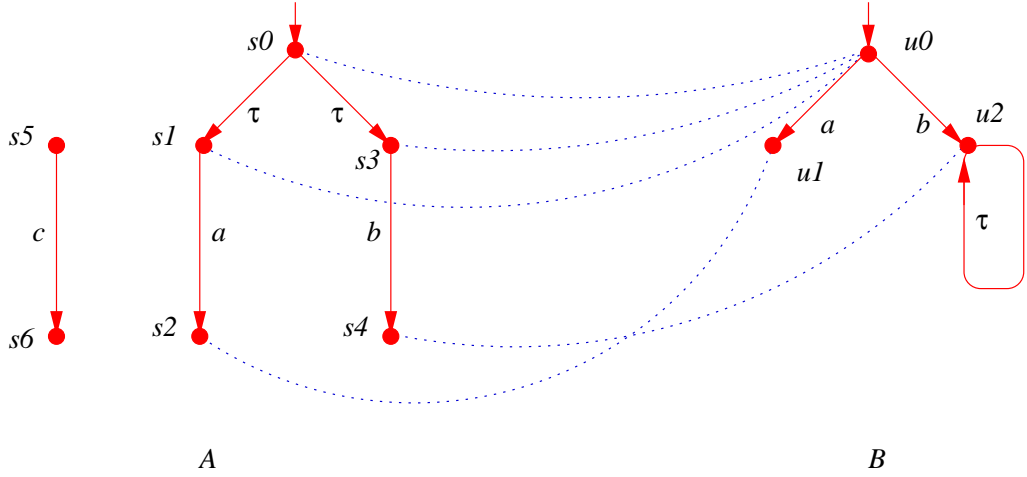}
\caption{\label{Fe:refinement}A step refinement.}
\end{center}
\end{figure}
Note that the $\tau$-steps in $A$ are not matched by any step in $B$.
Also the $c$-step in $A$ is not matched by any step in $B$: both source and
target states of this step are outside the domain of the step refinement.
This is allowed since both states are unreachable.
Observe that there is no step refinement from $B$ to $A$, but that there
exists a possibility mapping from $B$ to $A$.

Figure~\ref{Fe:anotherrefinement} gives another example.
In this case there is a step refinement from $A'$ to $B'$ but not from
$B'$ to $A'$.  There is not even a possibility mapping from $B'$ to $A'$.
\begin{figure}[ht]
\begin{center}
\epsfig{file=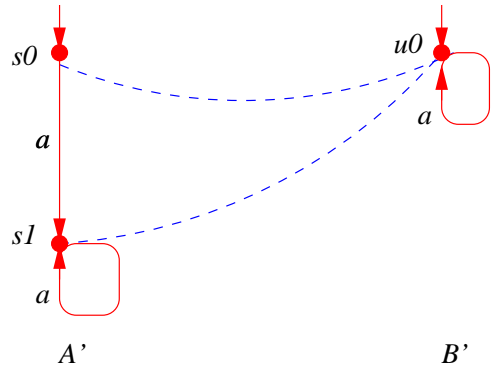}
\caption{\label{Fe:anotherrefinement}Another step refinement.}
\end{center}
\end{figure}
\end{example}

The following proposition states a basic sanity property of step refinements.

\begin{proposition}
\label{Pn:rcomposition}
$\prer$ is a preorder (i.e., is transitive and reflexive).
\end{proposition}
\begin{proof}
The identity function from $\states{A}$ to itself trivially is a step
refinement from $A$ to itself.  Hence $\prer$ is reflexive.
Transitivity follows from the observation that if $r$ is a step refinement
from $A$ to $B$ and $r'$ is a step refinement from $B$ to $C$, then the
function composition $r' \circ r$ is a step refinement from $A$ to $C$.
\end{proof}

\subsection{Execution Correspondence}
If there exists a step refinement from $A$ to $B$ then we can construct,
for each execution fragment of $A$,
a corresponding execution fragment of $B$ with the same trace.
The notion of `corresponding' is formalized below.

\begin{figure}[ht]
\begin{center}
\epsfig{file=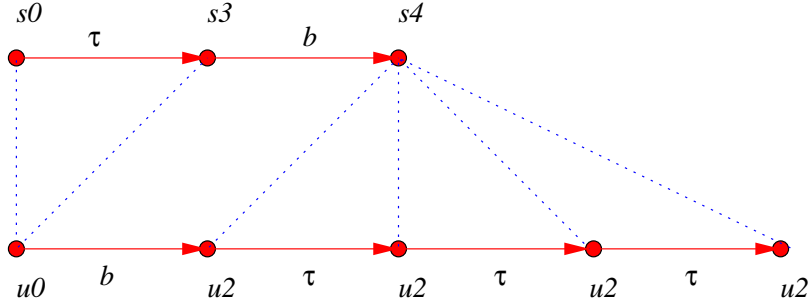}
\caption{\label{Fe:correspondence}Execution correspondence.}
\end{center}
\end{figure}

Suppose $A$ and $B$ are automata,
$R \subseteq \states{A} \times \states{B}$, and
$\alpha = s_0 a_1 s_1 a_2 s_2 \cdots$ and
$\alpha' = u_0 b_1 u_1 b_2 u_2 \cdots$ are execution fragments of $A$ and $B$,
respectively.
Let $\Index{\alpha}$ and $\Index{\alpha'}$ denote the index sets
of $\alpha$ and $\alpha'$.
Then $\alpha$ and $\alpha'$ {\em correspond via $R$} and are {\em $R$-related},
notation $(\alpha,\alpha') \in R$,
if there exists an {\em index relation} over $R$, i.e., a relation
$I \subseteq \Index{\alpha} \times \Index{\alpha'}$ such that
(1) if two indices are related by $I$ then the corresponding states are related
by $R$;
(2) $I$ is monotone;
(3) each index of $\alpha$ is related to an index of $\alpha'$ and vice versa;
(4) sides of ``squares'' always have the same label and
sides of ``triangles'' are labeled with $\tau$.
Formally we require,
for $i, i' \in \Index{\alpha}$ and $j, j' \in \Index{\alpha'}$,
\begin{enumerate}
\item
$(i, j) \in I ~~ \implies ~~ (s_i , u_j) \in R$
\item
$(i, j) \in I \wedge (i',j') \in I \wedge i < i' ~~ \implies ~~ j \leq j'$
\item
$I$ and $I^{-1}$ are total
\item
$\begin{array}[t]{lll}
(i, j) \in I \wedge (i+1,j+1) \in I & \implies & a_{i+1} = b_{j+1}\\
(i, j) \in I \wedge (i+1,j) \in I & \implies & a_{i+1} = \tau\\
(i, j) \in I \wedge (i,j+1) \in I & \implies & b_{j+1} = \tau
\end{array}$
\end{enumerate}
We write $(A, B) \in R$ if for every execution $\alpha$ of $A$ there
is an execution $\alpha'$ of $B$ such that $(\alpha,\alpha') \in R$, and
$[A, B] \in R$ if for every finite execution $\alpha$ of $A$
there is a finite execution $\alpha'$ of $B$ with $(\alpha,\alpha') \in R$.
%\begin{example}
Figure~\ref{Fe:correspondence} illustrates the correspondence between two
executions of automata $A$ and $B$ from Figure~\ref{Fe:refinement}.
%\end{example}

Another notion of correspondence has been presented by
Sogaard-Andersen, Lynch et al.\ \cite{GSSL93,SLL93} and
formalized by Mueller \cite{Mul98}.
Within the theory of I/O automata, execution correspondence plays a crucial
role in proofs of preservation of both safety and liveness properties.
Our notion is more restrictive than earlier work \cite{GSSL93,SLL93},
but technically simpler.
Moreover it has the advantage that it preserves `until' properties.
In this paper, we only study safety properties and it suffices to know that
corresponding executions have the same trace.
The latter fact is established in the next lemma.

\begin{lemma}
\label{lemma execution correspondence}
(Corresponding execution fragments have the same trace)
\begin{enumerate}
\item
Suppose $I$ is an index relation as above and $(i,j) \in I$.  Then
%\begin{eqnarray*}
$\trace{s_0 a_1 s_1 \cdots a_i s_i}  = \trace{u_0 b_1 u_1 \cdots b_j u_j}$.
%\end{eqnarray*}
\item
If $(\alpha,\alpha') \in R$ then $\trace{\alpha} = \trace{\alpha'}$.
\end{enumerate}
\end{lemma}
\begin{proof}
For (1), suppose $(i,j) \in I$.  By induction on $i+j$ we prove
\begin{eqnarray*}
\trace{s_0 a_1 s_1 \cdots a_i s_i} & = & \trace{u_0 b_1 u_1 \cdots b_j u_j}.
\end{eqnarray*}
If $i+j=0$ then both $i$ and $j$ are $0$.
Clearly, $\trace{s_0} = \trace{u_0} = \lambda$.

For the induction step, suppose $i+j> 0$.
For reasons of symmetry we may assume, without loss of generality, that $i> 0$.
Let $j'$ be the largest index with $j' \leq j$ and $(i-1,j') \in I$.
(By monotonicity, $i-1$ can only be related to indices less than or equal to
$j$, and by totality there is at least one such an index.)
We distinguish between three cases:
\begin{enumerate}
\item
$j' = j$.  Then by condition (4b), $a_i = \tau$.
By induction hypothesis,
\[
\trace{s_0 a_1 s_1 \cdots a_{i-1} s_{i-1}} =
\trace{u_0 b_1 u_1 \cdots b_j u_j}.
\]
Hence
$\trace{s_0 a_1 s_1 \cdots a_i s_i} = \trace{u_0 b_1 u_1 \cdots b_j u_j}$.
\item
$j' = j-1$.  Then by condition (4a), $a_i = b_j$.
By induction hypothesis,
\[
\trace{s_0 a_1 s_1 \cdots a_{i-1} s_{i-1}} =
\trace{u_0 b_1 u_1 \cdots b_{j-1} u_{j-1}}.
\]
Hence
$\trace{s_0 a_1 s_1 \cdots a_i s_i} = \trace{u_0 b_1 u_1 \cdots b_j u_j}$.
\item
$j' < j-1$.  Then by conditions (2) and (3), $(i, j-1) \in I$.
By condition (4c), this implies $b_j = \tau$.
By induction hypothesis,
\[
\trace{s_0 a_1 s_1 \cdots a_i s_i} =
\trace{u_0 b_1 u_1 \cdots b_{j-1} u_{j-1}}.
\]
Hence
$\trace{s_0 a_1 s_1 \cdots a_i s_i} = \trace{u_0 b_1 u_1 \cdots b_j u_j}$.
\end{enumerate}
This completes the proof of the induction step.

For (2), suppose that $(\alpha,\alpha') \in R$.
Then there exists an index relation $I$ that relates $\alpha$ and $\alpha'$.
Using (1) and the fact that both $I$ and $I^{-1}$ are total,
it follows that each finite prefix of
$\trace{\alpha}$ is also a finite prefix of $\trace{\alpha'}$, and vice versa.
This implies $\trace{\alpha} = \trace{\alpha'}$.
\end{proof}

The next corollary will be used repeatedly in the rest of this paper.
It states that in order to prove trace inclusion between automata $A$ and $B$
it suffices to find for each execution of $A$ a corresponding execution of
$B$.  Depending on whether one wants to prove inclusion of all traces or of
finite traces only, a stronger respectively  weaker type of
execution correspondence is required.

\begin{corollary}
\label{Cy:ect}
(Execution correspondence implies trace inclusion)
\begin{enumerate}
\item
If $(A, B) \in R$ then $[A,B] \in R$.
\item
If $[A, B] \in R$ then $A \preft B$.
\item
If $(A, B) \in R$ then $A \pret B$.
\end{enumerate}
\end{corollary}
\begin{proof}
Statement (1) follows from the definitions.
Statements (2) and (3) follow immediately from
Lemma~\ref{lemma execution correspondence} and the definitions.
\end{proof}

\subsection{Soundness and Partial Completeness}
The next theorem states that if there is a step refinement from $A$ to $B$,
it is possible to construct, for each execution of $A$, a corresponding
execution of $B$.
Using Corollary~\ref{Cy:ect}, this implies that step refinements
constitute a sound technique for proving trace inclusion.
In addition, the next theorem also allows us to use step refinements
as a sound technique for proving implementation relations between live
automata, as in previous work \cite{GSSL93,SLL93,Mul98}.

\begin{theorem}
\label{Tm:rsoundness}
(Soundness of step refinements)\\
If $r$ is a step refinement from $A$ to $B$ then $(A,B) \in r$.
\end{theorem}
\begin{proof}
Suppose $r$ is a step refinement from $A$ to $B$.
Let $\alpha = s_0 a_1 s_1 \cdots$ be an execution of $A$.
Inductively, we define an execution $\alpha' = u_0 b_1 u_1 \cdots$ of $B$
and an index relation $I$ such that $\alpha$ and $\alpha'$ are
$r$-related via $I$.

To start with, define $u_0 = r(s_0)$ and declare $(0,0)$ to be an
element of $I$.

Now suppose $(i,j) \in I$ and $i$ is a nonfinal index of $\alpha$.
We distinguish between two cases:
\begin{enumerate}
\item
If $r(s_i) \sarrow{a_{i+1}}{B} r(s_{i+1})$ then define
$b_{j+1} = a_{i+1}$, $u_{j+1} =  r(s_{i+1})$, and declare
$(i+1, j+1)$ to be an element of $I$;
\item
otherwise, declare $(i+1,j)$ to be an element of $I$.
\end{enumerate}
By construction, using the defining properties of a step refinement,
it follows that $I$ is an index relation.
This implies $(A,B) \in r$.
\end{proof}

Step refinements alone do not provide a complete method for proving
trace inclusion.  There is a partial completeness result, however.

\begin{theorem}
\label{Tm:rcompleteness}
(Partial completeness of step refinements)\\
Suppose $A$ is a forest, $B$ is deterministic and $A \preft B$.
Then $A \prer B$.
\end{theorem}
\begin{proof}
The relation $r \deq \after{B}\circ\past{A}$
is a step refinement from $A$ to $B$.
\end{proof}

Actually, we can even slightly strengthen the above theorem.  It suffices
to assume that $A$ restricted to its reachable states is a forest, and
that $B$ restricted to its reachable states is deterministic.
In Figure~\ref{Fe:refinement}, automaton $A$ restricted to its reachable
states is a forest and automaton $B$ is deterministic.  As we observed
already, there is a step refinement from $A$ to $B$.
Even if we restrict to reachable states, automaton $B$ is not
a forest and automaton $A$ is not deterministic.
As we observed, there is no step refinement from $B$ to $A$.

In practice, the preconditions of Theorem~\ref{Tm:rcompleteness} are
seldom met.
The higher-level specification often is deterministic, but it
rarely occurs that the lower-level specification is a forest.
Nevertheless, step refinements have been used in several substantial
case studies \cite{HSV94,NipS94,DGRV00}.

\section{Normed Forward Simulations}
\label{normed forward}
Even though there exists no step refinement from automaton $B'$ to automaton
$A'$ in Figure~\ref{Fe:anotherrefinement}, these automata do have the same
traces.
By moving from functions to relations it becomes possible to prove that
each trace of $B'$ is also a trace of $A'$.
This idea is formalized in the following definition.

A {\em normed forward simulation} from $A$ to $B$ consists of
a relation $f \subseteq \states{A}\times \states{B}$
and a function $n : \steps{A} \times \states{B} \rightarrow S$, for
some well-founded set $S$, such that
(here $f[s]$ denotes the set $\{ u \mid (s,u) \in f \}$):
\begin{enumerate}
\item
If $s \in \start{A}$ then $f[s] \cap \start{B} \neq \emptyset$.
\item
If $s \sarrow{a}{A} t \: \wedge \: u \in f[s]$ then
\begin{enumerate}
\item
$u \in f[t] \: \wedge \: a=\tau$, or
\item
$\exists v \in f[t] : u \sarrow{a}{B} v$, or
\item
$\exists v \in f[s] : u \sarrow{\tau}{B} v \: \wedge \:
n ( s \arrow{a} t, v ) < n ( s \arrow{a} t, u )$.
\end{enumerate}
\end{enumerate}
Write $A \pref B$ if there exists a normed forward simulation from
$A$ to $B$.

The intuition behind this definition is that if
$s \sarrow{a}{A} t$ and $(s,u) \in f$, then either
(a) the transition in $A$ is a stuttering step that does not have to be
matched, or
(b) there is a matching step in $B$, or
(c) $B$ can do a stuttering step which decreases the norm.
Since the norm decreases at each application of clause (c),
this clause can only be applied a finite number of times.
In general, the norm function may depend both on the transitions in $A$ and on
the states of $B$.  However, if $B$ is {\em convergent}, i.e., there are no
infinite $\tau$-paths, then one can simplify the type of the norm function
(though not necessarily the definition of the norm function itself) to
$n : \states{B} \rightarrow S$.  In fact, in the approach of
Groote and Springintveld \cite{GS95}, which not always applies to divergent
processes, the norm function is required to be of this restricted type.

\begin{example}
In Figure~\ref{Fe:anotherrefinement}, the relation indicated by the dashed
lines, together with an arbitrary norm function, is a normed forward
simulation from $B'$ to $A'$.

Consider automata $A$ and $B$ in Figure~\ref{Fe:refinement}.
Let $n$ be the function that assigns norm $1$ to state $s0$ and norm $0$
to all other states of $A$.
Then $n$ together with the relation indicated by the dashed lines
constitutes a normed forward simulation from $B$ to $A$.

\begin{figure}[ht]
\begin{center}
\epsfig{file=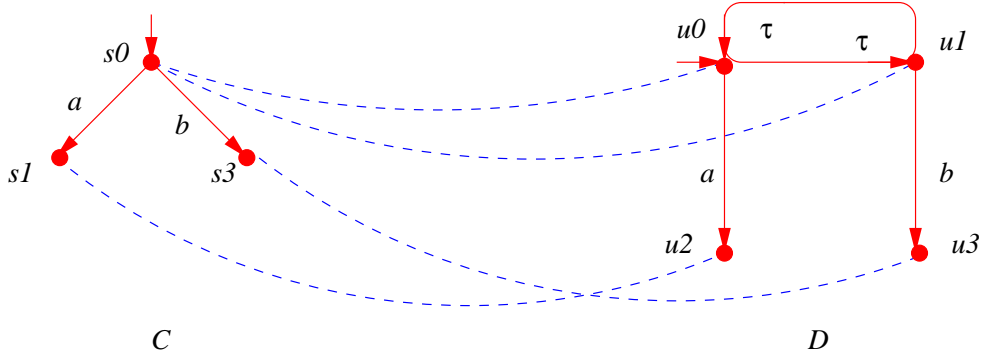}
\end{center}
\caption{Norm function must take steps of $C$ into account.}
\label{Fe:divergent}
\end{figure}
Now consider the automata $C$ and $D$ in Figure~\ref{Fe:divergent}.
Let $m$ be a norm function satisfying
\begin{eqnarray*}
m(s0 \arrow{a} s1, u0) ~ = ~  0 && m(s0 \arrow{a} s1, u1) ~ = ~ 1\\
m(s0 \arrow{b} s3, u0) ~ = ~ 1  && m(s0 \arrow{b} s3, u1) ~ = ~ 0
\end{eqnarray*}
Then $m$ together with the relation indicated by the dashed lines
constitutes a normed forward simulation from $C$ to $D$.
It is not hard to see that in this example, where $D$ is not convergent,
the norm necessarily depends on the selected step in $C$.

The example of Figure~\ref{Fe:divergent} also serves to illustrate
the difference between normed forward simulations and the forward
simulations that were studied by Jonsson \cite{Jo90,Jo91,Jo94}.
Essentially, Jonsson's forward simulations
are just normed forward simulations, except
that there is no norm function and condition 2(c) has been omitted.
We leave it to the reader to check that there exists no forward
simulation in this sense from $C$ to $D$.
This is the case even when we add ``stuttering'' $\tau$-loops
to each state, as required in Jonsson's models.
\end{example}

%Since each step refinement trivially induces a normed forward simulation,
%$A \prer B$ implies $A \pref B$.
The next proposition asserts that normed forward simulations indeed
generalize step refinements.

\begin{proposition}
\label{Pn:flifting}
$A \prer B$ $\implies$ $A \pref B$.
\end{proposition}
\begin{proof}
Together with an arbitrary norm function, any step refinement
(viewed as a relation) is a normed forward simulation.
\end{proof}

The soundness of normed forward simulations is trivially implied by the
following lemma and Corollary~\ref{Cy:ect}.

\begin{lemma}
\label{execution correspondence for forward simulations}
Suppose $(f, n)$ is a normed forward simulation from $A$ to $B$,
$A$ has an execution fragment $\alpha$ with first state $s$,
and $u$ is a state of $B$ with $u \in f[s]$.
Then $B$ has an execution fragment $\alpha'$ that starts in $u$
such that $(\alpha, \alpha') \in f$.
\end{lemma}
\begin{proof}
Let $c : \steps{A} \times \states{B}\rightarrow\{ L, C, R \} \times \states{B}$
be a function such that $c(s \arrow{a} t, u) = (x,v)$ and $u \in f[s]$ implies
\begin{enumerate}
\item
If $x = L$ then $u \in f[t] \: \wedge \: a=\tau$.
\item
If $x = C$ then $v \in f[t] \: \wedge \: u \sarrow{a}{B} v$.
\item
If $x = R$ then $v \in f[s] \: \wedge \: u \sarrow{\tau}{B} v \: \wedge \:
n ( s \arrow{a} t, v ) < n ( s \arrow{a} t, u )$.
\end{enumerate}
The existence of $c$, which chooses between a left move (L) of $A$,
a common move (C) of $A$ and $B$, or a right move (R) of $B$, is guaranteed
by the fact that $(f, n)$ is a normed forward simulation.

Let $\alpha = s_0 a_1 s_1 a_2 s_2 \cdots$.  Then $s = s_0$.
Inductively, we define a sequence $\sigma = z_0 z_1 z_2 \cdots$
of 4-tuples in $\nat \times \nat \times \acts{B} \times \states{B}$.
The first element in the sequence is $z_0 = (0,0,\tau,u)$.
If $z_k = (i,j,b,u)$ is an element of the sequence, and $i$ is a nonfinal
index of $\alpha$, then we define $z_{k+1}$ as follows
\begin{enumerate}
\item
If $c(s_i \arrow{a_{i+1}} s_{i+1}, u) = (L,v)$ then $z_{k+1} = (i+1,j,b,u)$.
\item
If $c(s_i \arrow{a_{i+1}} s_{i+1}, u) = (C,v)$ then
$z_{k+1} = (i+1,j+1,a_{i+1}, v)$.
\item
If $c(s_i \arrow{a_{i+1}} s_{i+1}, u) = (R,v)$ then
$z_{k+1} = (i,j+1, \tau, v)$.
\end{enumerate}
Suppose that both $(i,j,b,u)$ and $(i',j,b',u')$ occur in sequence $\sigma$.
We claim that $b = b'$ and $u = u'$.
To see why this is true assume without loss of generality that $(i,j,b,u)$
occurs before $(i',j,b',u')$.
Now observe that the values of both the first and second component of elements
in $\sigma$ increase monotonically.
This means that each successor of $(i,j,b,u)$ up to and including
$(i',j,b',u')$ has been obtained from its predecessor by applying rule (1).
This implies that the
the second respectively third components of all elements in the sequence from
$(i,j,b,u)$ until $(i',j,b',u')$ coincide.  Hence $b = b'$ and $u = u'$.

Using this property, we can define for each element $(i,j,b,u)$ in $\sigma$,
$b_j = b$ and $u_j = u$.
Let $\alpha' = u_0 b_1 u_1 b_2 u_2 \cdots$ and let
$I = \{ (i,j) \mid \exists b, u : (i,j,b,u) \mbox{ occurs in } \sigma \}$.
By construction of $\sigma$, using the properties of $c$, it follows that
$\alpha'$ is an execution fragment of $B$ that starts in $u$,
and that $I$ is an index relation over $f$.
This implies $(\alpha, \alpha') \in f$.
\end{proof}

\begin{theorem}
\label{Tm:fsoundness}
(Soundness of normed forward simulations)\\
If $f$ is a normed forward simulation from $A$ to $B$ then $(A,B) \in f$.
\end{theorem}
\begin{proof}
Immediate from the definitions and
Lemma~\ref{execution correspondence for forward simulations}.
\end{proof}

\begin{example}
Consider automata $C$ and $E$ in Figure~\ref{Fe:branching}.
\begin{figure}[ht]
\begin{center}
\epsfig{file=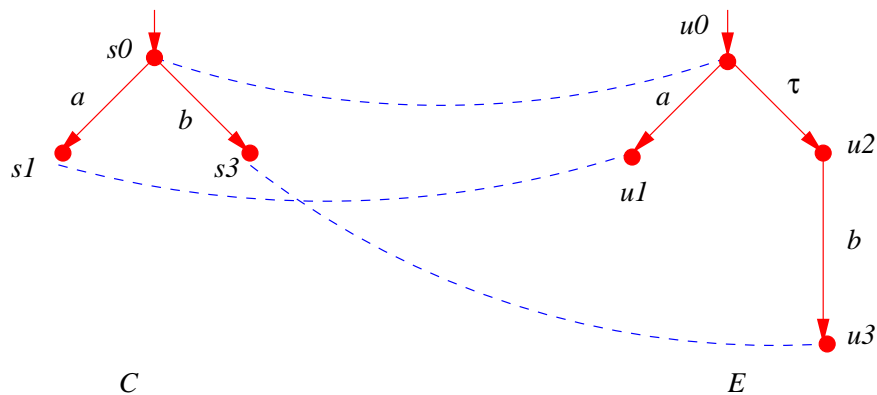}
\end{center}
\caption{Difference between forward simulations and normed forward simulations.}
\label{Fe:branching}
\end{figure}
There does not exist a normed forward simulation from $C$ to $E$.
Such a simulation would have to relate states $s0$ and $u0$.
But in order for $E$ to simulate the step $s0 \arrow{b} s3$,
it would also have to relates states $s0$ and $u2$.
But this is impossible since from state $u2$ there is no
way to simulate the step $s0 \arrow{a} s1$.

It turns out that there does exist a {\em forward simulation} in
Lynch and Vaandrager's sense \cite{LV95} from $C$ to $E$.
In the case of a forward simulation, a step of $A$ may be matched by a
sequence of steps in $B$ with the same trace.
This means that in the definition of a normed forward simulation
condition (2) is replaced by:
\begin{enumerate}
\item[2.]
If $s \sarrow{a}{A} t \:\wedge\: u \in f[s]$ then
$B$ has an execution fragment $\alpha$ with
$\first{\alpha} = u$, $\trace{\alpha} = \trace{a}$ and
$\last{\alpha} \in f[t]$.
\end{enumerate}
The dashed lines in Figure~\ref{Fe:branching} indicate a forward simulation
from $C$ to $E$.

The automata $A$ and $B$ in Figure~\ref{Fe:refinement} provide us with a
similar example: there exists a forward simulation from $B$ to $A$, but no
normed forward simulation.
\end{example}

The difference between forward simulations and normed forward simulations
is very similar to the difference between Milner's
{\em observation equivalence} \cite{Mi89} and the {\em branching bisimulation}
of Van Glabbeek and Weijland \cite{GW96}.
In fact, we can characterize normed forward simulations in terms of
``branching forward simulations'', a
notion that is inspired by the branching bisimulations of \cite{GW96}.
A similar characterization has been obtained by Namjoshi \cite{Nam97}
in the setting of stuttering bisimulations.

Formally, a {\em branching forward simulation} from $A$ to $B$ is a
relation $f \subseteq \states{A}\times \states{B}$ such that
\begin{enumerate}
\item
If $s \in \start{A}$ then $f[s] \cap \start{B} \neq \emptyset$.
\item
If $s \sarrow{a}{A} t$ and $u \in f[s]$ then
$B$ has an execution fragment that starts in $u$ and that is
$f$-related to $s \arrow{a} t$.
\end{enumerate}

The following theorem implies that there exists a normed forward simulation
between two automata if and only if there is a branching forward simulation
between them.

\begin{theorem}
\label{Tm:normed is branching}
\mbox{ }
\begin{enumerate}
\item
Suppose $(f,n)$ is a normed forward simulation from $A$ to $B$.
Then $f$ is a branching forward simulation from $A$ to $B$.
\item
Suppose $f$ is a branching forward simulation from $A$ to $B$.
Let $n(s \arrow{a} t, u)$ be $0$ if $u \not\in f[s]$
and otherwise be equal to the length of the shortest execution fragment
that starts in $u$ and that is $f$-related to $s \arrow{a} t$.
Then $(f,n)$ is a normed forward simulation from $A$ to $B$.
\end{enumerate}
\end{theorem}
\begin{proof}
Part (1) follows by
Lemma~\ref{execution correspondence for forward simulations}.
The proof of part (2) is routine.
\end{proof}

An interesting corollary of Theorem~\ref{Tm:normed is branching} is that
if there exists a normed forward simulation between two automata,
there is in fact a normed forward simulation with a norm that has the
natural numbers as its range.

\vspace{2ex}\par
The proof that branching bisimilarity is an equivalence is known to be
tricky \cite{Bas96}.  Likewise, the proof that branching forward simulations
induce a preorder is nontrivial.  We first need to define the auxiliary concept
of a {\em reduced} index relation and to prove a lemma about it.

Suppose that $\alpha$ and $\alpha'$ are $R$-related via index relation $I$.
We say that $I$ is {\em reduced} if the following two conditions are satisfied:
\begin{enumerate}
\item
If $\alpha$ is finite then $I$ relates the final index of $\alpha$
only to the final index of $\alpha'$.
\item
$I$ is {\em N-free}:
$(i,j) \in I \:\wedge\: (i+1,j+1) \in I ~~ \implies ~~
(i+1,j) \not\in I \:\wedge\: (i,j+1) \not\in I$.
\end{enumerate}
Observe that if $\alpha$ is finite and $I$ is reduced,
then $\alpha'$ is also finite.
The following technical lemma states that index relations can always be
reduced.

\begin{lemma}
\label{reduced}
Suppose that $\alpha$ and $\alpha'$ are $R$-related via index relation $I$.
Then $\alpha'$ has a prefix $\alpha''$ that is $R$-related to $\alpha$
via a reduced index relation $J \subseteq I$.
\end{lemma}
\begin{proof}
If $\alpha$ is infinite then let $\alpha'' = \alpha'$.
If $\alpha$ is finite then let
$\alpha''$ be the finite prefix of $\alpha'$ up to and including the first
state whose index is related by $I$ to the final index of $\alpha$.

Inductively we define a sequence $\sigma = z_0 z_1 z_2 \cdots$ of pairs
in $\nat\times\nat$.
The first element of the sequence is $z_0 = (0,0)$.
If $z_k = (i,j)$ is an element of the sequence and $i$ is a nonfinal index
then we define $z_{k+1}$ as follows:
\begin{enumerate}
\item
$(i+1,j+1) \in I ~~ \implies ~~ z_{k+1} = (i+1,j+1)$
\item
$(i+1,j) \in I \:\wedge\: (i+1,j+1) \not\in I ~~ \implies ~~ z_{k+1} = (i+1,j)$
\item
$(i,j+1) \in I \:\wedge\: (i+1,j+1) \not\in I ~~ \implies ~~ z_{k+1} = (i,j+1)$
\end{enumerate}
Note that since $I$ is an index relation, $z_{k+1}$ is properly defined.
Let $J = \{ (i,j) \mid (i,j) \mbox{ occurs in } \sigma \}$.
It is routine to check that $J \subseteq I$,
that $\alpha$ and $\alpha''$ are $R$-related via $J$,
and that $J$ is reduced.
A tricky point is the totality of $J$ and $J^{-1}$.
We prove that $J$ is total by contradiction.
Suppose that $J$ is not total.
Let $i$ be the smallest index of $\alpha$ with $J[i]=\emptyset$.
Let $j$ be the smallest index of $\alpha'$ with $(i,j) \in I$
($j$ exists since index relation $I$ is total).
Let $l$ be the maximal index of $\alpha'$ with $(i-1,l) \in J$
(there is a maximal index since $(i-1,l) \in J$ implies $(i-1,l) \in I$,
which implies $l \leq j$ by monotonicity of index relation $I$).
Let $z_k = (i-1,l)$.
Since $J[i]=\emptyset$, $z_{k+1} = (i-1,l+1)$.
Hence $(i-1,l+1) \in J$.
But this contradicts the fact that
$l$ be the maximal index of $\alpha'$ with $(i-1,l) \in J$.

In a similar way also the totality of $J^{-1}$ and N-freeness can
be proved by contradiction.
\end{proof}

We are now prepared to prove that branching forward simulations
(and hence also normed forward simulations) induce a preorder.

\begin{proposition}
\label{Pn:fcomposition}
$\pref$ is a preorder.
\end{proposition}
\begin{proof}
For reflexivity, observe that the identity function from $\states{A}$ to itself
is a branching forward simulation from $A$ to itself.

For transitivity, suppose $f$ and $g$ are branching forward simulations
from $A$ to $B$ and from $B$ to $C$, respectively.
We claim that $g \circ f$ is a branching forward simulation from $A$ to $C$.
It is trivial to check that $g \circ f$ satisfies condition (1) in the
definition of a branching forward simulation.  For condition (2),
suppose that $s \sarrow{a}{A} t \: \wedge \: u \in (g \circ f)[s]$.
Then there exists a state $w$ of $B$ such that $w \in f[s]$ and $u \in g[w]$.
Hence there is an execution fragment
$\alpha$ starting in $w$ such that $s \arrow{a} t$ and $\alpha$ are $f$-related
via some index relation $I$.
By Lemma~\ref{reduced}, we may assume that $I$ is reduced.
Also, there is an execution fragment
$\alpha'$ starting in $u$ such that $\alpha$ and $\alpha'$ are $g$-related via
some index relation $J$.
Again by Lemma~\ref{reduced}, we may assume that $J$ is reduced.
Using the fact that both $I$ and $J$ are reduced,
it is routine to check that $s \arrow{a} t$ and $\alpha'$ are
$g \circ f$-related via index relation $J \circ I$.
Thus $g \circ f$ satisfies condition (2) in the definition of a branching
forward simulation.
\end{proof}

Variants of the partial completeness result below appear in several papers
\cite{Jo87,LV95}.  Since higher-level specifications are often
deterministic, this result explains why in practice (normed) forward
simulations can so often be used to prove behavior inclusion.

\begin{theorem}
\label{Tm:fcompleteness}
(Partial completeness of normed/branching forward simulations)\\
If $B$ is deterministic and $A \preft B$ then $A \pref B$.
\end{theorem}
\begin{proof}
The relation $f \deq \after{B}\circ\past{A}$ is a
branching forward simulation from $A$ to $B$.
\end{proof}

It is interesting to note that there is one earlier result \cite{LV95}
concerning forward simulations that does not carry over to the normed/branching
simulations of this paper.
This result, Proposition 3.12, states that if $A$ is a forest
and $A \pref B$ then $A \prer B$.
The automata $C$ and $D$ of Figure~\ref{Fe:divergent} constitute a
counterexample.
Actually, the same Proposition 3.12 also does not carry over to the
setting of timed automata used earlier \cite{LV96}.

\section{Normed Backward Simulations}
\label{normed backward}
As we observed, there exists no normed forward simulation from automaton
$B$ to automaton $A$ in Figure~\ref{Fe:refinement}, even though both
automata have the same traces.
Also, there does not exist a normed forward simulation from automaton $C$ to
the trace equivalent automaton $E$ in Figure~\ref{Fe:branching}.
In both cases a forward simulation in Lynch and Vaandrager's
sense \cite{LV95} exists.
However, the example in Figure~\ref{Fe:backward} below shows that
also forward simulations do not yet provide us with a complete method for
proving trace inclusion.
It is well-known from the literature that completeness can be obtained by
adding some form of {\em backward simulation}.

\begin{example}
There exists no (normed/branching) forward simulation from automaton $C$
to automaton $F$ in Figure~\ref{Fe:backward}.
The relation indicated by the dashed lines fails since from state $u0$
the $b$-step from $s0$ can not be simulated, whereas from $u2$ the $a$-step
from $s0$ can not be simulated.
\begin{figure}[ht]
\begin{center}
\epsfig{file=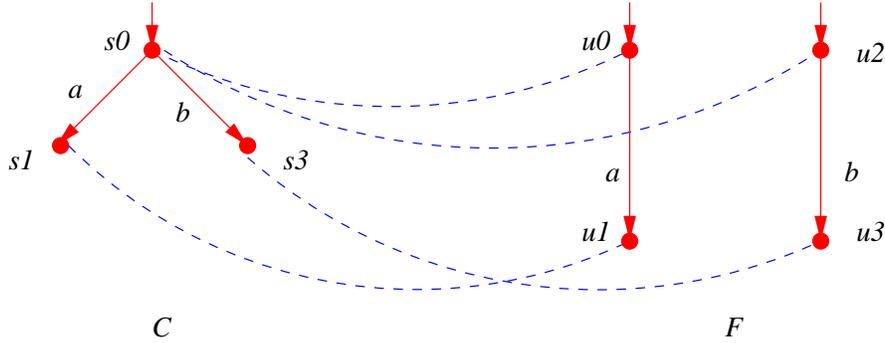}
\end{center}
\caption{The need for backward simulations.}
\label{Fe:backward}
\end{figure}
\end{example}

In many respects, backward simulations are the dual of forward simulations.
Whereas a forward simulation requires that {\em some} state in the
image of each start state should be a start state,
a backward simulation requires that {\em all}
states in the image of a start state be start states.
Also, a forward simulation requires that {\em forward} steps in the source
automaton can be simulated from related states in the target automaton,
whereas the corresponding condition for a backward simulations requires
that {\em backward} steps can be simulated.
However, the two notions are not completely dual:
the definition of a backward simulation contains a nonemptiness condition,
and also, in order to obtain soundness for general trace inclusion,
backward simulations also require a finite image condition.
The mismatch is due to the asymmetry in our automata between
the future and the past:
from any given state, all the possible histories are finite executions,
whereas the possible futures can be infinite.

Formally, we define
a {\em normed backward simulation} from $A$ to $B$ to be a pair of a total
relation $b \subseteq \states{A}\times \states{B}$ and a function
$n : (\steps{A} \cup \start{A}) \times \states{B} \rightarrow S$,
for some well-founded set $S$, satisfying
\begin{enumerate}
\item
If $s \in\start{A} \: \wedge \: u \in b[s]$ then
\begin{enumerate}
\item
$u \in \start{B}$, or
\item
$\exists v \in b[s] : v \sarrow{\tau}{B} u \:\wedge\: n(s,v) < n(s,u)$.
\end{enumerate}
\item
If $t \sarrow{a}{A} s \: \wedge \: u \in b[s]$ then
\begin{enumerate}
\item
$u \in b[t] \:\wedge\: a = \tau$, or
\item
$\exists v \in b[t] : v \sarrow{a}{B} u$, or
\item
$\exists v \in b[s] : v \sarrow{\tau}{B} u \:\wedge\:
n (t \arrow{a} s, v) < n (t \arrow{a} s, u)$.
\end{enumerate}
\end{enumerate}
Write $A \preb B$ if there is a normed backward simulation from $A$ to $B$,
and $A \preifb B$ if there is a normed backward simulation from $A$ to $B$ that
is image-finite.

\begin{example}
In Figure~\ref{Fe:backward}, the relation indicated by the dashed lines
is a normed backward simulation from $C$ to $E$, for arbitrary norm functions.
It is not difficult to construct normed backward simulations
from automaton $B$ to automaton $A$ in Figure~\ref{Fe:refinement}, and from
automaton $C$ to automaton $E$ in Figure~\ref{Fe:branching}.

\begin{figure}[ht]
\begin{center}
\epsfig{file=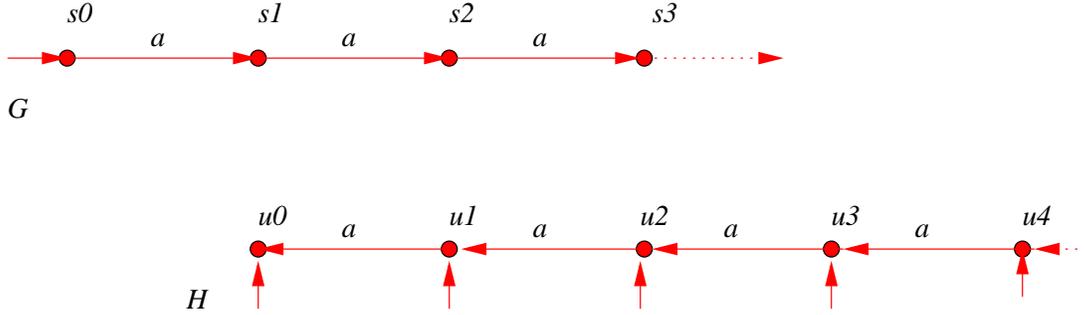}
\end{center}
\caption{No image-finite normed backward simulation.}
\label{Fe:fin}
\end{figure}
Figure~\ref{Fe:fin} illustrates the difference between $\preb$ and
$\preifb$.
Relation $\states{G}\times\states{H}$ together with an arbitrary norm
function constitutes a normed backward simulation from $G$ to $H$.
We claim that no image-finite normed backward simulation exist.
Because suppose that $b$ is such a relation.
Then, for all $i, j \in \nat$ with $i > 0$,
\begin{eqnarray*}
(si,uj) \in b & \implies & (si-1,uj+1) \in b
\end{eqnarray*}
This implies that
\begin{eqnarray*}
(si,uj) \in b & \implies & (s0,ui+j) \in b
\end{eqnarray*}
Since each state $si$ is related to at least one state $sj$, it follows
that state $s0$ is related to infinitely many states, which is a contradiction.
\end{example}

The following proposition states some trivial connections between
the preorders induced by normed backward simulations and step refinements.

\begin{proposition}
\label{Pn:blifting}
\mbox{ }
\begin{enumerate}
\item
If all states of $A$ are reachable and $A \prer B$ then $A \preifb B$.
\item
If $A \preifb B$ then $A \preb B$.
\end{enumerate}
\end{proposition}
\begin{proof}
Trivial.
\end{proof}

The next lemma is required to prove soundness of normed backward
simulations.

\begin{lemma}
\label{execution correspondence for backward simulations}
Suppose $(b, n)$ is a normed backward simulation from $A$ to $B$,
$A$ has a finite execution fragment $\alpha$ with last state $s$,
and $u$ is a state of $B$ with $u \in b[s]$.
Then $B$ has a finite execution fragment $\alpha'$ that ends in $u$
such that $(\alpha, \alpha') \in b$.
Moreover, if $\alpha$ is an execution then $\alpha'$ can be chosen
to be an execution as well.
\end{lemma}
\begin{proof}
Similar to the proof of
Lemma~\ref{execution correspondence for forward simulations}.
\end{proof}

By Lemma~\ref{execution correspondence for backward simulations}
and Corollary~\ref{Cy:ect}, the existence of a normed backward
simulation implies inclusion of finite traces.
Normed backward simulations, however, are in general not a sound method for
proving inclusion of infinite traces.  As a counterexample, consider automata
$G$ and $H$ from Figure~\ref{Fe:fin}.  There exists a normed backward
simulation from $G$ to $H$, but the infinite trace $a^{\omega}$ of $G$
is not a trace of $H$.
As is well-known from the literature, a sound method for proving inclusion
of infinite traces can be obtained by requiring image finiteness of the
simulation relation.

\begin{theorem}
\label{Tm:bsoundness}
(Soundness of normed backward simulations)
\begin{enumerate}
\item
If $b$ is a normed backward simulation from $A$ to $B$ then $[A,B] \in b$.
\item
If moreover $b$ is image-finite then $(A,B) \in b$.
\end{enumerate}
\end{theorem}
\begin{proof}
Statement (1) follows immediately by
Lemma~\ref{execution correspondence for backward simulations} and
the totality of $b$.
In order to prove (2), suppose that $b$ is image-finite.
Let $\alpha$ be an execution of $A$.
We have to establish the existence of an execution $\alpha'$ of $B$
with $(\alpha, \alpha') \in b$.
If $\alpha$ is finite then this follows by
Lemma~\ref{execution correspondence for backward simulations}
and the totality of $b$.
So assume that $\alpha$ is infinite.
We use a minor variation of K\H{o}nig's Lemma \cite{Kn97} presented by
Lynch and Vaandrager \cite{LV95}:
\begin{quote}
\em
Let $G$ be an infinite digraph such that
(1) $G$ has finitely many roots, i.e., nodes without incoming edges,
(2) each node of $G$ has finite outdegree, and
(3) each node of $G$ is reachable from some root.
Then there is an infinite path in $G$ starting from some root.
\end{quote}
The nodes of the graph $G$ that we consider are pairs $(I, \gamma)$
where $\gamma$ is a finite execution of $B$ and $I$ is an index
relation that relates $\gamma$ to some finite prefix of $\alpha$.
There is an edge from a node $(I, \gamma)$ to a node $(I', \gamma')$
iff $\gamma$ is a prefix of $\gamma'$ and $I'$ extends $I$ with
precisely one element.
It is straightforward to check that $G$ satisfies the conditions of
K\H{o}nig's Lemma.
Hence $G$ has an infinite path.
Let $J$ be the union of all the index relations occurring on nodes in
this path, and let
$\alpha'$ be the limit of the finite executions of the nodes in this path.
Observe that, by image-finiteness of $b$, each index of $\alpha$ occurs in
the domain of $J$.
Hence $(\alpha, \alpha') \in b$.
\end{proof}

The following Proposition~\ref{Pn:blifting2} is in a sense the converse of
Proposition~\ref{Pn:blifting}.  The proof is similar to that of the
corresponding result by Lynch and Vaandrager \cite{LV95}.

\begin{proposition}
\label{Pn:blifting2}
\mbox{ }
\begin{enumerate}
\item
If $B$ is deterministic and $A \preb B$ then $A \prer B$.
\item
If all states of $A$ are reachable, $B$ has fin
and $A \preb B$, then $A \preifb B$.
\end{enumerate}
\end{proposition}
\begin{proof}
For (1), suppose that $B$ is deterministic and that
$b$ is a normed backward simulation from $A$ to $B$.
Suppose that $s$ is a reachable state of $A$.
We will prove that $b[s]$ contains exactly one element.
Since any normed backward simulation that is functional on the reachable
states trivially induces a step refinement, this gives us $A \prer B$.

Because $b$ is a normed backward simulation it is a total relation, so we
know $b[s]$ contains at least one element.
Suppose that both $u_1 \in b[s]$ and $u_2 \in b[s]$; we prove $u_1 = u_2$.
Since $s$ is reachable, $A$ has an execution $\alpha$ that ends in $s$.
By Lemma~\ref{execution correspondence for backward simulations},
$B$ has executions $\alpha_1$ and $\alpha_2$ which end in $u_1$ and $u_2$,
respectively, such that $(\alpha,\alpha_1) \in b$ and
$(\alpha,\alpha_2) \in b$.
By Lemma~\ref{lemma execution correspondence},
$\trace{\alpha} = \trace{\alpha_1} = \trace{\alpha_2}$.
Now $u_1 = u_2$ follows by Lemma~\ref{La:beforeafter}(1), using the fact
the $B$ is deterministic.

For (2), suppose that all states of $A$ are reachable, $B$ has fin, and
$b$ is a normed backward simulation from $A$ to $B$.
Suppose that $s$ is a state of $A$.
Since $s$ is reachable, there is an execution $\alpha$ that ends in $s$.
Let $\beta$ be trace of $\alpha$.
By Lemma~\ref{execution correspondence for backward simulations} there exists,
for each $u \in b[s]$, an execution $\alpha_u$ of $B$ that ends in $u$
such that $(\alpha,\alpha_u) \in b$.
By Lemma~\ref{lemma execution correspondence}, $\trace{\alpha_u} = \beta$.
Hence $b[s] \subseteq \after{B}[\beta ]$.
But since $B$ has fin, $\after{B}[\beta ]$ is finite by
Lemma~\ref{La:beforeafter}(2).  Hence $b$ is image-finite.
\end{proof}

\begin{example}
Consider the two automata in Figure~\ref{Fe:backbranching}.
\begin{figure}[ht]
\begin{center}
\epsfig{file=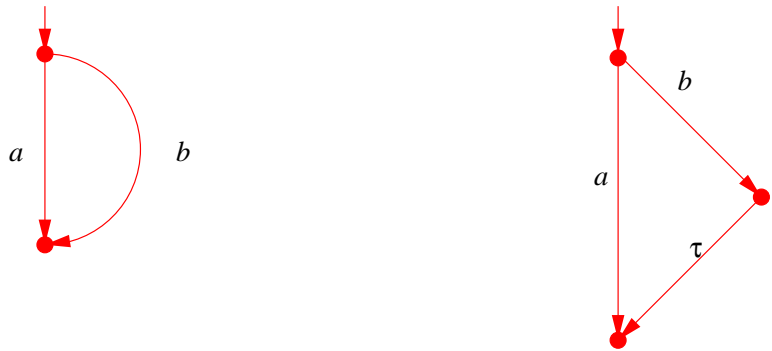}
\end{center}
\caption{Difference between backward simulations and normed backward
simulations.}
\label{Fe:backbranching}
\end{figure}
It is easy to see that there does not exist a normed backward
simulation from the first to the second automaton.
However, there does exist a {\em backward simulation} in
Lynch and Vaandrager's sense \cite{LV95}.
In such a backward simulation, a step of one automaton may be matched
by a sequence of steps in the other automaton with the same trace.
\end{example}

As in the forward case, we will now characterize normed backward simulations
in terms of ``branching backward simulations'', and use this characterization
to establish that $\preb$ and $\preifb$ are preorders.

A {\em branching backward simulation} from $A$ to $B$ is a total
relation $b \subseteq \states{A}\times \states{B}$ such that
\begin{enumerate}
\item
If $s \in \start{A}$ and $u \in b[s]$ then
$B$ has an execution that ends in $u$ and is $b$-related to $s$.
\item
If $t \sarrow{a}{A} s$ and $u \in f[s]$ then
$B$ has an execution fragment that ends in $u$ and is
$b$-related to $t \arrow{a} s$.
\end{enumerate}

\begin{theorem}
\label{Tm:normed backward is branching backward}
\mbox{ }
\begin{enumerate}
\item
Suppose $(b,n)$ is a normed backward simulation from $A$ to $B$.
Then $b$ is a branching backward simulation from $A$ to $B$.
\item
Suppose $b$ is a branching backward simulation from $A$ to $B$.
Let $n(s,u)$ be $0$ if $s$ is not a start state or $u \not\in b[s]$
and otherwise be equal to the length of the shortest execution that ends
in $u$ and is $b$-related to $s$.
Furthermore, let $n(t \arrow{a} s, u)$ be $0$ if $u \not\in f[s]$
and otherwise equal to the length of the shortest
execution fragment ending in $u$ that is $b$-related to $t \sarrow{a}{A} s$.
Then $(b,n)$ is a normed forward simulation from $A$ to $B$.
\end{enumerate}
\end{theorem}
\begin{proof}
Statement (1) follows by
Lemma~\ref{execution correspondence for backward simulations}.
The proof of statement (2) is routine.
\end{proof}

As in the forward case, we see that
if there exists a normed backward simulation between two automata,
there is in fact a normed backward simulation with a norm that has the
natural numbers as its range.

\begin{proposition}
\label{Pn:bcomposition}
$\preb$ and $\preifb$ are preorders.
\end{proposition}
\begin{proof}
Similar to the proof of Proposition~\ref{Pn:fcomposition}.
\end{proof}

The following partial completeness result is a variation of earlier results
\cite{Jo90,LV95}.

\begin{theorem}
\label{Tm:bcompleteness}
(Partial completeness of normed backward simulations)\\
If $A$ is a forest and $A \preft B$ then $A \preb B$.
\end{theorem}
\begin{proof}
The relation $b \deq \after{B}\circ\past{A}$ is a
branching backward simulation from $A$ to $B$.
\end{proof}

Note that by Proposition~\ref{Pn:blifting2} we can strengthen the
conclusion of Theorem~\ref{Tm:bcompleteness} to $A \preifb B$ in case
$B$ has finite invisible nondeterminism.

\begin{example}
Consider the automata $A'$ and $B'$ in Figure~\ref{Fe:anotherrefinement}.
There exists no normed backward simulation from $B'$ to $A'$.
The relation indicated by the dashed lines fails since the backward
transition from state $u0$ cannot be simulated from the related state $s0$.
Consequently, normed backward simulations do not provide a complete
proof method for establishing trace inclusion.
In the next section, we will see that completeness can be obtained by
combining normed forward and backward simulations.
\end{example}

\section{Normed History Relations}
\label{Sn:history relations}
In this section we define {\em normed history relations}.
These provide an abstract view of the {\em history variables}
of Abadi and Lamport \cite{AL91}, which in turn are abstractions of
the {\em auxiliary variables} of Owicki and Gries \cite{OG76}.

A pair $(r,n)$ is a {\em normed history relation} from $A$ to $B$
if $r$ is a step refinement from $B$ to $A$, and
$(\inverse{r},n)$ is a normed forward simulation from $A$ to $B$.
Write $A \preh B$ if there exists a normed history relation from $A$ to $B$.

Clearly $A \preh B$ implies $A \pref B$ and $B \prer A$.
Through these implications, the preorder and soundness results for normed
forward simulations and step refinements carry over to normed history relations.
In fact, if $(r,n)$ is a normed history relation from $A$ to $B$ then $r$
is just a functional {\em branching bisimulation} from $B$ to $A$ in the
sense of Van Glabbeek and Weijland \cite{GW96}.
Hence, history relations preserve behavior of automata in a very strong sense.
Intuitively, there is a history relation from $A$ to $B$ if $B$ can be
obtained from $A$ by adding an extra state variable that records
information about the history of an execution.

\begin{example}
Consider again the automata $A'$ and $B'$ in Figure~\ref{Fe:anotherrefinement}.
Together with an arbitrary norm function, the dashed lines constitute a normed
history relation from $B'$ to $A'$.
Because, as we observed, there is no step refinement from $B'$ to $A'$,
there exists no normed history relation from $A'$ to $B'$.
\end{example}

An important example of a history relation is provided by the ``unfolding''
construction.
The {\em unfolding} of an automaton $A$, notation $\unfold{A}$, is the
automaton obtained from $A$ by recording the complete history of an
execution.  Formally, $\unfold{A}$ is the automaton $B$ defined by
\begin{itemize}
\item
$\states{B}= \fexecs{A}$,
\item
$\start{B} =$ the set of executions of $A$ that consist of a single start state,
\item
$\acts{B} = \acts{A}$, and
\item
for $\alpha', \alpha\in\states{B}$ and $a \in\acts{B}$,
$\alpha'\sarrow{a}{B}\alpha ~~\Leftrightarrow ~~
\alpha=\alpha' \: a \: \last{\alpha}$.
\end{itemize}
The next proposition relates an automaton to its unfolding.

\begin{proposition}
\label{Pn:unfold}
\label{Pn:unfoldhistory}
$\unfold{A}$ is a forest and $A \preh \unfold{A}$.
\end{proposition}
\begin{proof}
Clearly, $\unfold{A}$ is a forest.
The function ${\it last}$ which maps each finite execution
of $A$ to its last state is a step refinement from $\unfold{A}$ to $A$,
and the relation $\inverse{{\it last}}$, together with an arbitrary norm
function, is a normed forward simulation from $A$ to $\unfold{A}$.
\end{proof}

The following completeness theorem, a variation of a result due to Sistla
\cite{Sis91}, asserts that normed history relations together with normed
backward
simulations constitute a complete proof method for establishing trace inclusion.
Consequently, also normed forward simulations together with normed backward
simulations constitute a complete proof method.

\begin{theorem}
\label{Tm:Sistla}
(Completeness of normed history relations and normed backward simulations)\\
If $A \preft B$ then there exists an automaton $C$
such that $A \preh C \preb B$.
\end{theorem}
\begin{proof}
Take $C = \unfold{A}$.
By Proposition~\ref{Pn:unfoldhistory}, $C$ is a forest and $A \preh C$.
Since $A \preft B$, also $C \preft B$ by soundness of
history relations.
Next apply the partial completeness result for backward simulations
(Theorem~\ref{Tm:bcompleteness}) to conclude $C \preb B$.
\end{proof}

Observe that if we can assume in addition that $B$ has fin, we may replace
$\preb$ by $\preifb$ in the conclusion using Proposition~\ref{Pn:blifting2}.

Normed forward simulations are equivalent to normed history variables combined
with step refinements: whenever there is a normed forward simulation from
$A$ to $B$, we can find an intermediate automaton $C$ such that there is a
normed history relation from $A$ to $C$ and a step refinement from $C$ to $B$.
The converse implication trivially holds since normed history relations and
step refinements are special cases of normed forward simulations.
In order to prove the existence of automaton $C$, we need to define a notion of
``superposition'' of automata and to prove a technical lemma.

Let $R \subseteq \states{A}\times \states{B}$ be a relation with
$R \cap (\start{A} \times \start{B} ) \neq \emptyset$.
The {\em superposition} $\superp{A}{R}{B}$ of $A$ and $B$ via $R$ is the
automaton $C$ defined by
\begin{itemize}
\item
$\states{C}= R$,
\item
$\start{C}= R \cap ( \start{A} \times \start{B} )$,
\item
$\acts{C}=\acts{A}\cap\acts{B}$, and
\item
for $(s, u), (t, v) \in\states{C}$ and $a \in\acts{C}$,
$(s, u) \sarrow{a}{C} (t, v) ~~ \Leftrightarrow$
\begin{eqnarray*}
& & a = \tau \wedge s = t \wedge u \sarrow{\tau}{B} v\\
&\vee  & a = \tau \wedge u = v \wedge s \sarrow{\tau}{A} t\\
&\vee & s \sarrow{a}{A} t \wedge u \sarrow{a}{B} v.
\end{eqnarray*}
\end{itemize}
Essentially, the superposition $\superp{A}{R}{B}$ is just the usual parallel
composition of $A$ and $B$ with the set of states restricted to $R$.

\begin{lemma}
\label{La:fsuperposition}
Suppose $(f,n)$ is a normed forward simulation from $A$ to $B$.
Let $C = \superp{A}{f}{B}$ and let $\pi_1$ and $\pi_2$ be the projection
functions that map states of $C$ to their first and second components,
respectively.
Let $n'$ be the norm function given by $n' (\delta,u) = n (\delta,\pi_2(u))$.
Then $(\pi_1,n')$ is a normed history relation from $A$ to $C$, and
$\pi_2$ is a step refinement from $C$ to $B$.
\end{lemma}
\begin{proof}
Straightforward from the definitions.
\end{proof}

\begin{theorem}
\label{Tm:forwardvshistory}
$A \pref B$ $\Iff$ $( \exists C : A \preh C \prer B)$.
\end{theorem}
\begin{proof}
Forward implication follows by Lemma~\ref{La:fsuperposition}.
For backward implication, suppose $A \preh C \prer B$.
Then $A \pref C$ by the definition of history relations, and
$C \pref B$ because any step refinement is a normed forward simulation.
Now $A \pref B$ follows by the fact that $\pref$ is a preorder.
\end{proof}

Klop and Ariola \cite{AriK96}[Intermezzo 3.23] state a remarkable result:
on a domain of of finitely branching process graphs (i.e., automata
considered modulo isomorphism)
the preorder induced by functional bisimulations (i.e., history relations) is
in fact a partial order: $A \preh B$ and $B \preh A$ implies $A = B$.
They also present a counterexample to show that the
finite branching property is needed to prove this result.
Below we present a slight generalization of their result \cite{AriK96} in
the setting of our paper.
It turns out to be sufficient to assume that automata have finite invisible
nondeterminism (fin).

\begin{theorem}
\label{Tm:history po}
Suppose $A$ and $B$ have fin, $A \preh B$ and $B \preh A$.
Then the reachable subautomata of $A$ and $B$ are isomorphic.
\end{theorem}
\begin{proof}
Suppose that $(f,n)$ is a normed history relation from $A$ to $B$,
and $(g,m)$ is a normed history relation from $B$ to $A$.
Because $A$ and $B$ have fin, both $\start{A}$ and $\start{B}$ are finite.
Since $f$ is a step refinement, it maps start states of $B$ to start
states of $A$.  Using the fact that $f^{-1}$ is a forward simulation,
we infer that $f$ is surjective on start states.
Hence $\mid \start{B} \mid ~ \leq ~ \mid \start{A} \mid$.
By a similar argument, using the fact that $(g,m)$ is a normed history
relation from $B$ to $A$, we obtain
$\mid \start{A} \mid ~ \leq ~ \mid \start{B} \mid$.
This means that $f$ is also injective on start states.

Let $\beta$ be an arbitrary trace of $A$ and $B$.  Using a similar
argument as above, we infer
\begin{eqnarray*}
f ( \after{A}[\beta] ) & = & \after{B}[\beta]\\
g ( \after{B}[\beta] ) & = & \after{A}[\beta]
\end{eqnarray*}
Since, by Lemma~\ref{La:beforeafter}(2),
both $\after{A}[\beta]$ and $\after{B}[\beta]$ are finite, it follows that
\begin{eqnarray*}
\mid \after{A}[\beta] \mid & = & \mid \after{B}[\beta] \mid
\end{eqnarray*}
This means that $f$ and $g$ are injective on the sets
$\after{B}[\beta]$ and $\after{A}[\beta]$, respectively.
But since each reachable state is in a set $\after{B}[\beta]$ or
$\after{A}[\beta]$, for some $\beta$, it follows that $f$ and $g$
are injective on all states.
Now the required isomorphism property follows from the fact that
$f$ and $g$ are step refinements.
\end{proof}

Intuitively, one may interpret the above result as follows: if $A \preh B$
then $B$ contains as much {\em history information} as $A$.
If $B$ contains as much history information as $A$, and $A$ contains as much
history information as $B$, then they are equal.

\section{Normed Prophecy Relations}

In this section, we will define normed prophecy relations and show that they
correspond to normed backward simulations, very similarly to the
way in which normed history relations correspond to normed forward simulations.

A pair $(r,n)$ is a {\em normed prophecy relation} from $A$ to $B$
if $r$ is a step refinement from $B$ to $A$ and
$(\inverse{r},n)$ is a normed backward simulation from $A$ to $B$.
We write $A \prep B$ if there is a normed prophecy relation from $A$ to $B$,
and $A \preifp B$ if there is a normed prophecy relation $(r,n)$
with $\inverse{r}$ image-finite.
Thus $A \preifp B$ implies $A \preifb B$ and $A \prep B$,
and $A \prep B$ implies $A \preb B$ and $B \prer A$.
Moreover, if all states of $A$ are reachable, $B$ has finite invisible
nondeterminism and $A \prep B$, then $A \preifp B$.
It is easy to check that the preorder and soundness results for backward
simulations and refinements carry over to prophecy relations.

The following lemma is the analogue of Lemma~\ref{La:fsuperposition} in
the backward setting.  Using this lemma, we can prove that normed backward
simulations are equivalent to normed prophecy variables combined with
step refinements.

\begin{lemma}
\label{La:bsuperposition}
Suppose $(b,n)$ is a normed backward simulation from $A$ to $B$.
Let $C =\superp{A}{b}{B}$ and
let $\pi_1$ and $\pi_2$ be the projection functions that map states of $C$
to their first and second components, respectively.
Let $n'$ be the norm function given by $n' (\delta,u) = n (\delta,\pi_2(u))$.
Then $(\pi_1, n')$ is a normed prophecy relation from $A$ to $C$, and
$\pi_2$ is a step refinement from $C$ to $B$.
If $b$ is image-finite then so is $\inverse{\pi_1}$.
\end{lemma}

\begin{theorem}
\label{Tm:backwardvsprophecy}
\mbox{ }
\begin{enumerate}
\item
$A \preb B$ $\Iff$ $( \exists C : A \prep C \prer B)$.
\item
$A \preifb B$ $\Iff$ $( \exists C : A \preifp C \prer B)$.
\end{enumerate}
\end{theorem}
\begin{proof}
Analogous to that of Theorem~\ref{Tm:forwardvshistory},
using Lemma~\ref{La:bsuperposition}.
\end{proof}

We can now prove variants of the well-known completeness result of
Abadi and Lamport \cite{AL91}.

\begin{theorem}
\label{Tm:AbadiLamport}
(Completeness of normed history+prophecy relations and step refinements)\\
Suppose $A \preft B$.  Then
\begin{enumerate}
\item
$\exists C, D : A \preh C \prep D \prer B$.
\item
If $B$ has fin then $\exists C, D : A \preh C \preifp D \prer B$.
\end{enumerate}
\end{theorem}
\begin{proof}
By Theorem~\ref{Tm:Sistla},
there exists an automaton $C$ with $A \preh C \preb B$.
Next, Theorem~\ref{Tm:backwardvsprophecy} yields the required automaton
$D$ with $C \prep D \prer B$, which proves (1).
The proof of (2) is similar, but uses Proposition~\ref{Pn:blifting2}.
\end{proof}

The following theorem states that $\prep$ is a partial order on the
class of automata with fin, considered modulo isomorphism of
reachable subautomata.
The proof is analogous to that of Theorem~\ref{Tm:history po}, the
corresponding result for normed history relations.

\begin{theorem}
\label{Tm:prohecy po}
Suppose $A$ and $B$ have fin, $A \prep B$ and $B \prep A$.
Then the reachable subautomata of $A$ and $B$ are isomorphic.
\end{theorem}

\section{Decidability}
Thus far, our exposition has been purely semantic.
In the words of Abadi and Lamport \cite{AL91}:
``We have considered specifications, but not the languages
in which they are expressed.  We proved the existence of refinement mappings,
but said nothing about whether they are expressible in any language.''
In this section, we move to the syntactic world and discuss some
decidability issues.  To this end we have to fix a language for defining
automata.  The language below can be viewed as a simplified version of
the IOA language of Garland et al.\  \cite{IOA-LANG}.

We assume an underlying assertion language $\cal L$ which is a first-order
language over interpreted symbols for expressing functions and predicates
over some concrete domains such as integers, arrays, and lists of integers.
If $X$ is a set of (typed) variables then we write $F(X)$ and
$E(X)$ for the collection of formulas and expressions, respectively,
in which variables from $X$ may occur free.
An automaton can be described syntactically by first specifying a finite
set $X$ of variables, referred to as the {\em state variables}.
For each state variable $x$ we assume the presence of a copy $x'$, called the
{\em primed version} of $x$.
We write $X'$ for the set $\{ x' \mid x \in X \}$ and, if $\phi$ is a formula
then we write $\phi'$ for the formula obtained from $\phi$ by replacing
each occurrence of a state variable by its primed version.
The set of states of the automaton is defined as the set of all
valuations of the state variables in $X$.
The set of initial states is specified by a predicate in $F(X)$,
called the {\em initial condition}.
The actions are specified via a finite number of {\em action names} with,
for each action name $a$, a finite list $\vec{v}$ of variables
called the {\em parameters} of $a$.
We assume $\{ \vec{v} \} \cap X = \emptyset$.
The set of actions of the automaton is defined as the union,
for each action name $a$, of all tuples $a (\vec{d})$, where
$\vec{d}$ is a valuation of the parameters $\vec{v}$ in their respective
domains.
The transition relation is specified by providing, for each action name $a$
with parameters $\vec{v}$, a {\em transition predicate} in
$F(X \cup \{ \vec{v} \} \cup X')$, i.e., a predicate that
may contain action parameters as well as primed and unprimed state variables.

\begin{example}
\label{Channel}
Below we specify a FIFO channel in IOA syntax \cite{IOA-LANG}.
\begin{verbatim}
automaton Channel
  states
    buffer: Seq[Nat]
  initial condition
    buffer = {}
  actions
    send(v: Nat),
    receive(v: Nat),
    tau
  transitions
    action send(v)
      predicate buffer' = buffer |- v
    action receive(v)
      predicate buffer ~= {} /\ v = head(buffer)
                  /\ buffer' = tail(buffer)
    action tau
      predicate false
\end{verbatim}
In IOA datatypes are specified using the Larch specification
language \cite{GH93}.  In the example we use the standard finite list
datatype, with {\tt \{\}} denoting the empty list, 
{\tt |-} denotes the opereration that appends an element to the
end of a list, etc.
Transitions are specified in a standard predicative style.
The example automaton has no $\tau$ transitions, which is
specified by the transition predicate {\tt false}.

This piece of syntax defines an automaton $A$ with
\begin{itemize}
\item
$\states{A} = \nat^\ast$,
\item
$\start{A} = \{ \lambda \}$,
\item
$\acts{A} = \{ {\tt send}(d),~ {\tt receive}(d) \mid d \in\nat\}\cup\{\tau\}$,
\item
$\steps{A}$ is the least set that contains the following elements, for all
$\sigma \in \nat^\ast$ and $d \in \nat$,
\begin{eqnarray*}
\sigma & \arrow{{\tt send}(d)} & \sigma \: d\\
d \: \sigma & \arrow{{\tt receive}(d)} & \sigma.
\end{eqnarray*}
\end{itemize}
\end{example}

Now assume that we have specified two automata $A$ and $B$, using
state variables $\vec{x}$ and $\vec{y}$, respectively.
Let $X = \{ \vec{x} \}$ and $Y = \{ \vec{y} \}$.
Assume $X \cap Y = \emptyset$.

A step refinement from $A$ to $B$ can be specified by a formula
of the form $\theta \wedge \vec{y} = \vec{e}$,
with $\theta \in E(X)$ and $\vec{e}$ a list of expressions in $E(X)$
that matches $\vec{y}$ in terms of length and types.
In this formula, the first conjunct defines the domain of the step refinement
whereas the second conjunct defines a map from states of $A$ to states of $B$
by specifying, for each state variable of $B$, its value in terms of the
values of the state variables of $A$.

A normed forward simulation can be described by a predicate
in $F(X \cup Y)$ together with, for each action type
$a$ with parameters $\vec{v}$, an expression in
$E(X \cup \{ \vec{v} \} \cup X' \cup Y)$ that
specifies the norm function.
In practice, norm functions often only depend on the states of $B$, which means
that they can be specified by means of a single expression in $E(Y)$.

\begin{example}
\label{TwoChannels}
Consider the following specification, essentially just the
chaining of two FIFO channels.
\begin{verbatim}
automaton TwoChannels
  states
    buffer1: Seq[Nat],
    buffer2: Seq[Nat]
  initial condition
    buffer1 = {} /\ buffer2 = {}
  actions
    send(v: Nat),
    receive(v: Nat),
    tau
  transitions
    action send(v)
      predicate buffer1' = buffer1 |- v /\ buffer2' = buffer2
    action receive(v)
      predicate buffer2 ~= {} /\ v = head(buffer2)
                  /\ buffer2' = tail(buffer2) /\ buffer1' = buffer1
    action tau
      predicate buffer1 ~= {} /\ buffer1' = tail(buffer1) /\
                 /\ buffer2' = buffer2 |- head(buffer1)
\end{verbatim}
Let $B$ be the automaton denoted by this specification.  It is easy to
prove that the formula below (where {\tt ||} denotes concatenation of lists)
defines a step refinement from $B$ to the
automaton $A$ of Example~\ref{Channel}.
\begin{verbatim}
        buffer = buffer2 || buffer1
\end{verbatim}
It is also routine to check that this formula together with the norm
on states of $B$ defined by
\begin{verbatim}
        if buffer1 ~= {} /\ buffer2 = {} then 1 else 0
\end{verbatim}
defines a normed forward simulation from $A$ to $B$.
\end{example}

We will now show that, under some reasonable (sufficient but certainly
not necessary) assumptions, it is in fact
decidable whether a given predicate/expression indeed corresponds to
a step refinement or normed forward simulation.
Assume that automaton $A$ is described using state variables $\vec{x}$,
initial condition $\varphi_0$ and, for each action name $a$,
a transition predicate $\varphi_a$.
Likewise, assume that automaton $B$ is described using state variables
$\vec{y}$, initial condition $\psi_0$ and, for each action name $a$,
a transition predicate $\psi_a$.
Assume further that each action name $a$ of $A$ is also an action name of $B$,
and that $a$ has the same parameters in both $A$ and $B$.
Write $P_a$ for the list of parameters of $a$.
We require that $P_\tau = \emptyset$.

Suppose that we want to check whether a formula
$\rho \deq \theta \wedge \vec{y} = \vec{e}$ denotes a step refinement.
This is equivalent to proving validity of the following formula:
\begin{eqnarray*}
&        & \varphi_0 ~~ \implies ~~ \theta\\
& \bigwedge & \varphi_0 \wedge \rho ~~ \implies ~~ \psi_0\\
& \bigwedge_a & \varphi_a \wedge \theta  ~~ \implies ~~ \theta'\\
& \bigwedge_{a \neq \tau} & \varphi_a \wedge \rho \wedge \rho'  ~~ \implies ~~
		\psi_a\\
& \bigwedge & \varphi_\tau \wedge \rho \wedge \rho'  ~~ \implies ~~
		\psi_\tau \vee \vec{y}=\vec{y'}
\end{eqnarray*}
In this formula, the first conjunct asserts that the function is
defined for start states of $A$; the second conjunct that
start states of $A$ are mapped onto start states of $B$; the third
conjunct that if the function is defined for the source
of a transition then it is also defined for the target state of a
transition; and the two final conjuncts encode the transfer condition.
Thus checking whether a partial function is a step refinement from $A$ to $B$
is decidable if the partial function as well as $A$ and $B$ can all be
expressed within a fragment of $\cal L$ for which tautology checking is
decidable.

Next suppose that we want to check whether a formula $\rho$ together with
norm expressions $n_a$, for each action name $a$, denotes a normed forward
simulation from $A$ to $B$.
In order to turn this into a decidable question, we have to make some
additional assumptions about the specification of $B$.
We assume that $B$ has finitely many start states\footnote{This assumption
can be relaxed if we assume that the value of certain state variables of
$B$ is fully determined by $\rho$ and the state of $A$:
for those state variables the initial value can be left unspecified.},
which are listed explicitly,
i.e., we require that the initial condition $\psi_0$ is of the form
\begin{eqnarray}
\label{finitestart}
\psi_0 & = & \bigvee_{i \in I_0} \vec{y} = \vec{e_0^i}
\end{eqnarray}
where $I_0$ is a finite index set
and, for each $i$, $ \vec{e_0^i}$ is a list of closed terms.
In addition we assume that in any state and for any given value of the action
parameters, only finitely many transitions are possible in $B$, which are
listed explicitly.
Formally we require that, for each action type $a$,
transition predicate $\psi_a$ is of the form
\begin{eqnarray}
\label{finitetransitions}
\psi_a & = & \bigvee_{i \in I_a} (\chi_a^i \wedge \vec{y'} = \vec{e_a^i})
\end{eqnarray}
where $I_a$ is a finite index set and, for each $i$,
$\chi_a^i$ is a formula in $F(Y \cup \{ P_a \})$ and $\vec{e_0^i}$ is a
list of expressions in $E(Y \cup \{ P_a \})$.
Basically, $\chi_a^i$ gives the precondition of the $i$-th instance of
transition $a$ and $\vec{y'} = \vec{e_a^i}$ specifies the effect of taking it.
Both assumption (\ref{finitestart}) and (\ref{finitetransitions}) are
satisfied by most automaton specifications that one encounters in practice.
In particular, the assumptions hold for the channels specified in
Examples \ref{Channel} and \ref{Channel}.
Only specifications that involve a nondeterministic choice that is not
a priori bounded fall outside of our format.  An example of this, described
by Sogaard-Andersen et al.\ \cite{SGGLP93},
is a FIFO channel in which a crash action may result in
the loss of an arbitrary subset of the messages contained in a buffer.
Under assumptions (\ref{finitestart}) and (\ref{finitetransitions}), we
can eliminate the existential quantifiers that occur in the definition of
a normed forward simulation, and checking the conditions in this definition
becomes equivalent to proving validity of the following formula:
\begin{eqnarray*}
& & \varphi_0 ~~ \implies ~~ \bigvee_{i \in I_0} \rho[\vec{e_0^i}/\vec{y}]\\
& \bigwedge_{a \neq \tau} & \varphi_a \wedge \rho ~~ \implies ~~
\bigvee_{i \in I_a} (\chi_a^i \wedge \rho'[\vec{e_a^i}/\vec{y'}]) \vee
\bigvee_{i \in I_\tau} (\chi_\tau^i \wedge \rho[\vec{e_\tau^i}/\vec{y}]
	\wedge n_a [\vec{e_\tau^i}/\vec{y}] < n_a )\\
& \bigwedge & \varphi_\tau \wedge \rho ~~ \implies ~~
\rho'[\vec{y}/\vec{y'}] \vee
\bigvee_{i \in I_\tau} (\chi_\tau^i \wedge \rho'[\vec{e_\tau^i}/\vec{y'}]) \vee
\bigvee_{i \in I_\tau} (\chi_\tau^i \wedge \rho[\vec{e_\tau^i}/\vec{y}]
	\wedge n_\tau [\vec{e_\tau^i}/\vec{y}] < n_\tau )
\end{eqnarray*}
If this formula can be expressed within a fragment of $\cal L$
for which tautology checking is decidable
then it is decidable whether $\rho$ together with expressions $n_a$
constitutes a normed forward simulation.
It is easy to see that a similar result can also be obtained for
normed history variables.
Thus far, however, we have not been able to come up with plausible
syntactic restrictions, applicable in practical cases,
that ensure decidability of normed backward
simulations and/or normed prophecy relations.
It is for instance not clear how one can eliminate the existential
quantifier in the formula that asserts that in a normed backward simulation
for each state of $A$ there exists a related state of $B$.
We think this constitutes an interesting area for future research.

Our decidability results for step refinements and normed forward simulations do
not carry over to the refinements and forward simulations as described,
for instance, by Lynch and Vaandrager \cite{LV95}.
In order to see this, let $A$ be a system with two states, an initial and
a final one, and a single transition labeled {\it halt} from the initial
to the final state.
Let $B$ be a system that simulates the $n$-th Turing machine such that
each computation step of the Turing machine corresponds with a $\tau$-move,
and that moves via a {\it halt}-action to a designated final state if and only
if the computation of the Turing machine terminates.
The function that maps the initial state of $A$ to the initial state of $B$
and the final state of $A$ to the final state of $B$ is a weak refinement
iff the $n$-th Turing machine halts.
It is straightforward to specify $A$, $B$ and the function from states of
$A$ to states of $B$ in a decidable logic.
Hence it is undecidable whether a given function is a weak refinement,
even in a setting where the underlying logic is decidable.

\section{Reachability}

For the sake of simplicity, all definitions of simulations and
refinements so far have been presented without any mention of
reachability or invariants.
However, in practical verifications it is almost always the case that
first some invariants (properties that hold for all reachable states)
are established for the lower-level and/or higher-level specification.
These invariants are then used in proving the step correspondence.
In this section we show how to integrate reachability concerns
into the simulation definitions.
More specifically, we present adapted versions of step refinements,
normed forward simulations and normed backward simulations which
include reachability concerns, and discuss their relationship with
the original definitions.
For examples of the use of these adapted definitions and their formalization
in PVS, we refer to our earlier work \cite{Gri00}.

An adapted {\em step refinement} from $A$ to $B$ consists of
a partial function $r : \states{A}\rightarrow\states{B}$
satisfying the following two conditions:
\begin{enumerate}
\item
If $s \in \start{A}$ then $s\in\domain{r}$ and $r(s) \in \start{B}$.
\item
If $s \sarrow{a}{A} t \:\wedge\: s\in\domain{r} \:\wedge\: {\it reachable}(A,s)
\:\wedge\: {\it reachable}(B,r(s))$ then $t\in\domain{r}$ and
\begin{enumerate}
\item
$r(s)=r(t) \: \wedge \: a = \tau$, or
\item
$r(s) \sarrow{a}{B} r(t)$.
\end{enumerate}
\end{enumerate}
Clause ${\it reachable}(A,s)$ in condition (2) allows us to reuse
invariants that have previously been established for lower-level
specification $A$, whereas clause ${\it reachable}(B,r(s))$ in condition (2)
makes it possible to reuse known invariants of higher-level specification $B$.
The adapted definition can easily be seen as a special case of
the original definition in Section~\ref{Dn:StepRefinements}:
if $r$ is an adapted step refinement then the restriction $r'$ of $r$
defined by
\begin{eqnarray*}
s \in\domain{r'} & \deq &
s\in\domain{r}\:\wedge\:{\it reachable}(A, s)\:\wedge\:{\it reachable}(B, r(s)),
\end{eqnarray*}
is a regular step refinement.
Conversely, any regular step refinement trivially satisfies the conditions
of the adapted version.

An adapted {\em normed forward simulation} from $A$ to $B$ consists of
a relation $f \subseteq \states{A}\times \states{B}$
and a function $n : \steps{A} \times \states{B} \rightarrow S$, for
some well-founded set $S$, such that:
\begin{enumerate}
\item
If $s \in \start{A}$ then $f[s] \cap \start{B} \neq \emptyset$.
\item
If $s \sarrow{a}{A} t \: \wedge \: u \in f[s]
\:\wedge\: {\it reachable}(A,s) \:\wedge\: {\it reachable}(B,u)$ then
\begin{enumerate}
\item
$u \in f[t] \: \wedge \: a=\tau$, or
\item
$\exists v \in f[t] : u \sarrow{a}{B} v$, or
\item
$\exists v \in f[s] : u \sarrow{\tau}{B} v \: \wedge \:
n ( s \arrow{a} t, v ) < n ( s \arrow{a} t, u )$.
\end{enumerate}
\end{enumerate}
Again, the clause ${\it reachable}(A,s)$ in condition (2) allows us to reuse
invariants that have previously been established for $A$,
whereas clause ${\it reachable}(B,u)$ in condition (2)
permits reuse of invariants of $B$.
And again the adapted definition can easily been seen as a special
case of the original definition (in Section~\ref{normed forward}):
if $(f,n)$ is an adapted normed forward simulation then the pair
$(g,n)$, where $g = f \cap ({\it reachable}(A) \times {\it reachable}(B))$,
is a regular normed forward simulation.
Conversely, any regular normed forward simulation trivially is an
adapted normed forward simulation.

An adapted {\em normed backward simulation} from $A$ to $B$ consists
of a relation $b \subseteq \states{A}\times \states{B}$, a predicate
$Q \subseteq \states{B}$, and a function
$n : (\steps{A} \cup \start{A}) \times \states{B} \rightarrow S$,
for some well-founded set $S$, such that:
\begin{enumerate}
\item
If $s \in\start{A} \: \wedge \: u \in b[s] \: \wedge \: Q(u) \: $ then
\begin{enumerate}
\item
$u \in \start{B}$, or
\item
$\exists v \in b[s]: v \sarrow{\tau}{B} u \:\wedge\: n(s,v) < n(s,u) \:\wedge\:
Q(v)$.
\end{enumerate}
\item
If $t \sarrow{a}{A} s \:\wedge\: u \in b[s] \:\wedge\: {\it reachable}(A,t)
\:\wedge\: Q(u)$ then
\begin{enumerate}
\item
$u \in b[t] \:\wedge\: a = \tau$, or
\item
$\exists v \in b[t] : v \sarrow{a}{B} u \:\wedge\: Q(v)$, or
\item
$\exists v \in b[s]: v \sarrow{\tau}{B} u \:\wedge\:
n(t \arrow{a} s, v) < n(t \arrow{a} s, u) \:\wedge\: Q(v)$.
\end{enumerate}
\item
If ${\it reachable}(A,s)$ then $\exists u \in b[s] : Q(u)$.
\end{enumerate}
Clause ${\it reachable}(A,t)$ in condition (2) allows us to reuse
invariants that have previously been established for $A$,
and clause $Q(u)$ in condition (2) permits reuse of invariants of $B$.
Note that by a trivial inductive argument a backward simulation
can never relate a reachable state of $A$ to a non-reachable state of $B$.
Thus we can safely restrict the range of any backward simulation by all
invariants proven for $B$.  To this end predicate $Q$ has been included
in the definition of the adapted normed backward simulation, even though
strictly speaking
(1) $Q$ need not be an invariant, and
(2) $Q$ can always be eliminated by restricting the range of $b$.
Once more the adapted definition is a special case of the original definition
(in Section~\ref{normed backward}):
if $(b,n)$ is an adapted normed backward simulation then $(b,n)$ is also
a regular normed backward simulation from the automaton $A'$,
that restricts $A$ to its reachable states, to the automaton $B'$,
that restricts $B$ to the states in $Q$.
Conversely, any regular normed backward simulation trivially is an
adapted normed backward simulation with $Q = \states{B}$.

We leave it up to the reader to work out adapted versions of the
normed history and prophecy relations.